\documentclass[acmsmall,screen,nonacm]{acmart}

\usepackage[linesnumbered,ruled,vlined]{algorithm2e}

\usepackage{subcaption}

\usepackage{tabularx,multirow,graphicx}
\usepackage[export]{adjustbox}
\usepackage{textcomp}
\usepackage{xcolor}
\usepackage{xspace}
\usepackage{setspace}
\usepackage{underscore}

\usepackage{hyperref}
\hypersetup{colorlinks=true,allcolors=blue}
\usepackage[skip=3pt]{caption} 

\usepackage[inline]{enumitem}
\usepackage{enumitem}

\usepackage{listing}
\usepackage{listings}
\usepackage{xstring}
\newcommand{\chisel}[1]{%
  \StrChar{#1}{1}[\firstchar]%
  \IfStrEqCase{\firstchar}{%
    {A}{\chiselBlue{#1}}%
    {B}{\chiselBlue{#1}}%
    {C}{\chiselBlue{#1}}%
    {D}{\chiselBlue{#1}}%
    {E}{\chiselBlue{#1}}%
    {F}{\chiselBlue{#1}}%
    {G}{\chiselBlue{#1}}%
    {H}{\chiselBlue{#1}}%
    {I}{\chiselBlue{#1}}%
    {J}{\chiselBlue{#1}}%
    {K}{\chiselBlue{#1}}%
    {L}{\chiselBlue{#1}}%
    {M}{\chiselBlue{#1}}%
    {N}{\chiselBlue{#1}}%
    {O}{\chiselBlue{#1}}%
    {P}{\chiselBlue{#1}}%
    {Q}{\chiselBlue{#1}}%
    {R}{\chiselBlue{#1}}%
    {S}{\chiselBlue{#1}}%
    {T}{\chiselBlue{#1}}%
    {U}{\chiselBlue{#1}}%
    {V}{\chiselBlue{#1}}%
    {W}{\chiselBlue{#1}}%
    {X}{\chiselBlue{#1}}%
    {Y}{\chiselBlue{#1}}%
    {Z}{\chiselBlue{#1}}%
  }[\chiselBlack{#1}]%
}
\newcommand{\chiselBlue}[1]{%
  {\fontsize{0.92em}{1em}\bfseries\ttfamily\color{blue}#1}%
}
\newcommand{\chiselBlack}[1]{%
  {\fontsize{0.92em}{1em}\ttfamily #1}%
}
 \newcommand{\verilog}[1]{\chisel{#1}}

\usepackage{tikz}

\usepackage{eso-pic}

\usepackage{booktabs}

\usepackage{units} 

\newcommand{\Gbps}[1]{\unit[#1]{Gbps}}

\newcommand{\MHz}[1]{\unit[#1]{MHz}}

\acmJournal{TRETS}
\acmVolume{0}
\acmNumber{0}
\acmArticle{0}
\acmMonth{00}

\begin{document}

\title[PAF: Pipeline Automation Framework]{Pipeline Automation Framework for Reusable High-throughput Network Applications on FPGA}

\author{Jean Bruant}
\orcid{0000-0001-8151-0392} 
\affiliation{%
  \institution{OVHcloud; Univ. Grenoble Alpes, CNRS, Grenoble INP$^1$, TIMA}
  \country{France}
}

\author{Pierre-Henri Horrein}
\orcid{0000-0002-9714-7871}
\email{pierre-henri.horrein@ovhcloud.com}
\affiliation{%
  \institution{OVHcloud}
  \city{Lyon}
  \country{France}
}

\author{Olivier Muller}
\orcid{0000-0002-4182-0502}
\email{olivier.muller@univ-grenoble-alpes.fr}
\affiliation{%
  \institution{Univ. Grenoble Alpes, CNRS, Grenoble INP$^1$, TIMA}
  \streetaddress{46 avenue Felix Viallet}
  \postcode{38031}
  \city{Grenoble}
  \country{France}
}

\author{Fr\'{e}d\'{e}ric P\'{e}trot}
\orcid{0000-0003-0624-7373}
\email{frederic.petrot@univ-grenoble-alpes.fr}
\affiliation{%
  \institution{Univ. Grenoble Alpes, CNRS, Grenoble INP$^1$, TIMA}
  \city{Grenoble}
  \country{France}
}

\begin{abstract}
In a context of ever-growing worldwide communication traffic, cloud service providers aim at deploying scalable infrastructures to address heterogeneous needs.
Part of the network infrastructure, FPGAs are tailored to guarantee low-latency and high-throughput packet processing.
However, slowness of the hardware design process impairs FPGA ability to be part of an agile infrastructure under constant evolution, from incident response to long-term transformation.
Deploying and maintaining network functionalities across a wide variety of FPGAs raises the need to fine-tune hardware designs for several FPGA targets.
To address this issue, we introduce \emph{PAF}, an architectural parameterization framework based on a pipeline-oriented design methodology.
\emph{PAF (Pipeline Automation Framework)} implementation is based on \emph{Chisel}, a Scala-embedded Hardware Construction Language (HCL), that we leverage to interface with circuit elaboration.
Applied to industrial network packet classification systems, \emph{PAF} demonstrates efficient parameterization abilities, enabling to reuse and optimize the same pipelined design on several FPGAs.
In addition, \emph{PAF} focuses the pipeline description on the architectural intent, incidentally reducing the number of lines of code to express complex functionalities.
Finally, \emph{PAF} confirms that automation does not imply any loss of tight control on the architecture by achieving on par performance and resource usage with equivalent exhaustively described implementations.

\end{abstract}
\footnotetext[1]{Institute of Engineering Univ. Grenoble Alpes}



\maketitle

\section{Introduction}

\begin{figure}[bp]
  \begin{center}
    \includegraphics[width=0.90\textwidth]{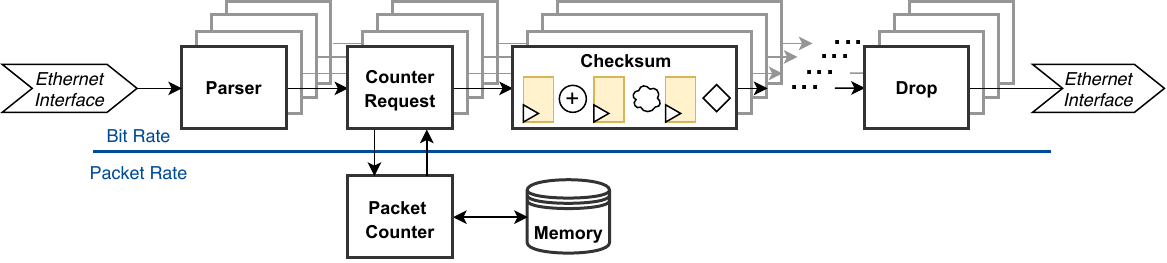}
    \caption{Network application pipelined representation}
    \label{fig:intro:netapp}
  \end{center}
\end{figure}

Cloud infrastructures and their underlying network infrastructures are constantly evolving to address various needs, from incident response to scalability to maintenance.
As reconfigurable hardware accelerators, FPGAs are able to provide both high-throughput and low-latency required by networking algorithms, while also exhibiting fast update capabilities.
With several millions of configurable logic cells available in the largest FPGAs, taking full advantage of these massively parallel resources is a design challenge.
In demanding network application context, fine-grained pipeline architectures are designed to address traffic rate requirements, with fine-tuning of each processing block of the overall network operation.
These blocks are hierarchically composed both in sequence---data dependency requirements---and in parallel---throughput requirements---drawing higher-level pipelines, up to an applicative macro-pipeline level as Figure~\ref{fig:intro:netapp} illustrates.
This figure interestingly exhibits another characteristic of network pipelines: some operations such as checksum computation are done at bitrate, while others such as packet filtering or state storage are done at packet rate.
These coexisting representations of the network flow lead to strong synchronization needs between the main pipeline at bitrate and the packet rate operations, for example based on \emph{start of packet} or \emph{end of packet} control signals.

Based on such high-level applicative representation, traditional hardware development process then consists in describing the applicative algorithms as functionally equivalent hardware architectures.
These hardware descriptions are not only responsible for the functional intent, but also for many fine-grained implementation details associated to a given hardware target.
Descriptions concentrate multiple concerns in a single place, inherited from both functional and implementation constraints.
In particular, pipeline implementations with traditional hardware description languages closely tie functional intent to both local and global signals synchronization.
The impact of a local modification on the whole system might become complex to follow and validate, which fundamentally impairs the ability of the design to follow the pace of required evolutions.
This intricate link also slows down designers during development phases, when the design undergoes many iterations from early drafts to deliverable state.

Functional parameterization -- i.e. the ability to parameterize the design depending on functionality intents -- can help to smoothen evolutions and iterations.
However, its usefulness remains strictly limited to anticipated scopes, and it requires costly engineering efforts.
Moreover, functional parameterization, as opposed to architectural parameterization, does not primarily intent to address the ability to target multiple FPGA devices but rather adapt the overall functionality to fit a given target.

The distinction between functional parameterization and architectural -- i.e. non-functional -- parameterization is highly dependent of the context.
Widths of signals in fixed-point arithmetic are functional parameters as they directly impact the quality of result.
The number of register stages to achieve such -- highly regular -- computation is purely architectural. 
It results of a target-dependent trade-off between combinational complexity and resource usage to meet timing requirements. 
In packet—processing context, computation is not a regular processing of a continuous flow of samples.
It is event-based, with every single packet requiring dedicated operations, independently (or not) of the size of the packet.
In particular, most processing steps cannot be determined before collecting the result of the previous steps.
The core logic mostly consists of control and synchronization operations with low combinational complexity.
In this context, we aim at introducing architectural parameters, with the following objectives:
\begin{itemize}
  \item abstract away error-prone signal synchronization operations,
  \item automate the insertion of configurable architectural elements,
  \item allow an architectural parameterization tailored to each FPGA target, 
  \item avoid cluttering the description with verbose parameterization,
  \item maintain a tight control on the architecture performance.
\end{itemize}

A successful architectural parameterization should be a zero-cost abstraction in terms of quality of result and should not impair the readability of the functionality.
In this paper, we present how pipeline-oriented hardware descriptions associated to configurable signal synchronization strategies can provide such architectural parameterization.
Our contributions are the following:
\begin{itemize}
   \item a design methodology, based on signal synchronization and non-exhaustive descriptions,
   \item a graph-based hardware signal synchronization model with several resolution strategies,
   \item a framework implementing the design methodology,
   \item an industrial use case evaluating the quality of the resulting circuits in terms of latency, throughput and resource usage against previous implementations.
\end{itemize}

The rest of the paper is organized as follows:
Section~\ref{sec:related:work} compares our approach to literature.
Section~\ref{sec:architectural:param} introduces two architectural parameters relevant to pipelined hardware descriptions.
Sections~\ref{sec:framework:implementation} and~\ref{sec:sync:reso} detail the algorithms and their implementation in our \emph{Pipeline Automation Framework (PAF)}.
Section~\ref{sec:xp:indus} details an industrial packet-processing application and the architectural parameterization achieved with \emph{PAF} in this context.
Section~\ref{sec:xp:fifo:explo} further details the automation and analysis abilities provided by the framework, to help designers select architectural parameters.
Finally, section~\ref{sec:conclusion} concludes this paper and details future works.

\section{Related Work}
\label{sec:related:work}

The pipelining paradigm is a recognized approach to take advantage of the parallelism in digital circuits.
The first pipeline-oriented descriptions and patterns have been introduced in the early history of digital design~\cite{cotten1965circuit}\cite{hallin1972pipelining}.
Such patterns allow to maximize the throughput and hardware resource usage.  
Numerous researches have proposed abstraction models for hardware pipelines and automated tools to generate complex and configurable architectures.

In practice, pipelining consists in splitting a functionality in successive stages with intermediate registers.
This operation, often referred as \emph{scheduling}, remains a manual process in traditional Hardware Description Languages (HDLs), but its automation is an active research field.
High-Level Synthesis (HLS)---in which software languages are synthesized into hardware architectures~\cite{canis2011legup}\cite{choi2016unified}---typically aims at inferring the most efficient scheduling from high-level algorithm expressions.
Various approaches have been explored, in particular for dataflow~\cite{marinescu2001specs} and signal processing~\cite{icstoan2017automating}.
Recent works on dynamic scheduling of dataflow circuits~\cite{josipovic2018dynamically}, based on the theory of latency-insensitive designs~\cite{carloni2001theory}, target higher circuit performance by inserting buffers on critical backpressure paths~\cite{josipovic2021buffer}.
Similar approaches have also proven to bring flexibility to microarchitecture design~\cite{galceran2010automatic}.

By focusing on functionality and algorithmic expressions, these technics offer a high automation degree to designers, in exchange for a large delegation of control over the resulting architecture to the tools.
Some parameterization of the generated architecture remains accessible through the use of \emph{pragmas} which are generally needed to achieve the best results~\cite{cheng2022finding}.
This parameterization also improves the reusability of designs, with limited code modifications to explore various architectures.
However, in practice \emph{pragmas} are highly dependent of each FPGA vendor and its associated toolchain, hence requiring considerable efforts to reuse the same design across multiple targets.

To keep full control of the architecture and its performance, manual scheduling remains required in many cases, which explains---despite their obsolescence---the indisputable popularity of traditional HDLs such as \emph{SystemVerilog} and \emph{VHDL}.
To overcome the limitations of traditional HDLs, most notably in terms of verbosity, expressiveness and parameterization, numerous alternatives have been proposed recently.

Among them, Hardware Construction Languages (HCLs), such as Chisel~\cite{bachrach2012chisel} or PyMTL~\cite{lockhart2014pymtl} leverage modern software paradigms to introduce advanced configuration and generation capabilities.
Instead of describing circuits, HCLs focus on describing hardware generators and provide the ability to build custom complex abstractions on top of relatively basic hardware description capabilities. 
Such generators can be highly parameterized, but HCLs do not provide any architectural parameterization by design and additional frameworks must be developed to offer such features. 
The \emph{DFiant} hardware construction language~\cite{port2017dfiant} integrates pipeline scheduling strategies inherited from HLS, with similar features and limitations.
Several other formalization efforts and abstractions can be compared to our work.
A SystemC library~\cite{harcourt2014systemc} introduces a Domain Specific Language (DSL) to describe pipelines with transactional-level modeling (TLM), based on event-driven description.
It provides the ability to instantiate both parallel and sequential pipelines stages, unlocking fast exploration of scheduling technics on a design.
Similarly, TL-Verilog~\cite{hoover2017timing} provides a stage-based description, enabling easier identification and iteration over pipeline stages, with the ability to implement a stage as register or wire.
The rust-based HDL \emph{Filament}~\cite{nigam2023modular} goes further by providing a compile-time checking of static pipeline scheduling.
While bringing high expressiveness for pipeline-oriented description, these approaches do not to tackle the issue of architectural parameterization.
They do not aim either at automating the synchronization of signals within the pipeline.

From a network perspective, several domain-specific researches have emerged to improve the flexibility of network pipeline descriptions.
FlowBlaze~\cite{pontarelli2019flowblaze} is a DSL that provides a way to represent network pipelines with a focus on stateful applications support.
Another popular language is P4~\cite{bosshart2014p4}, which relies on a clear separation of concerns.
On the one hand, an architecture representation defines the fine-grained capabilities of processing elements.
On the other hand, the overall application is represented as the successive operations to be applied on each network packets.
Network application development based on P4 toolchains bring actual improvements over traditional HDL implementation, such as native support of standard protocols, reducing time required for both design and debug.
While P4 increases expressiveness and reusability in network context, it suffers from several limitations, such as relying on vendor-dependent implementations, defeating the reusability purpose across multiple targets~\cite{wang2017p4fpga}.
Moreover, it does not provide any pipeline-oriented hardware abstraction at a fine-grained scale or architectural parameterization.
In practice, the base hardware functions must be implemented using the same abovementioned hardware description methodologies, hence struggling with the exact same limitations.

Compared to the literature, our approach aims at introducing an intermediate path between the automated scheduling of HLS and the exhaustive verbosity of traditional HDLs and HCLs.
We aim at leveraging the high-level programming paradigms and the reflexivity over the circuit under elaboration brought by HCLs to expose some architectural parameters.
Exposing these parameters requires to first build a proper model of the pipeline scheduling.
However, we do not intend to automate pipeline scheduling based on data dependency analysis, but rather allow  designers to describe their pipelines with a clear expression of their own scheduling choices.
Such architectural parameters could then become a new field of exploration for HLS tools and their Design Space Exploration (DSE) algorithms, in the quest of finding further optimizations in an extended design space.
To our knowledge, this is the first approach to offer an explicit user-defined pipeline scheduling model used to provide architectural parameterization towards improved design reusability.
Moreover, our approach does not aim at creating a standalone DSL taking ownership of entire designs but rather an automation to be integrated only when relevant within existing Chisel hardware designs.
Finally, both the pipeline model and synchronization resolution efforts are fully agnostic of the source language and equivalent pipeline automation frameworks could be adapted for virtually any HDL or HCL.

The next sections detail how the reusability requirements naturally lead to architectural parameterization and the need to decouple these parameters from pipeline scheduling and functionality description.

\section{Architectural Parameterization of Pipelines}
\label{sec:architectural:param}
\subsection{Motivations}
\label{sec:dec:context}
\begin{figure}[ht]
  \begin{center}
    \vspace*{-0.5em}
    \includegraphics[trim=43 495 45 90,clip, width=\textwidth]{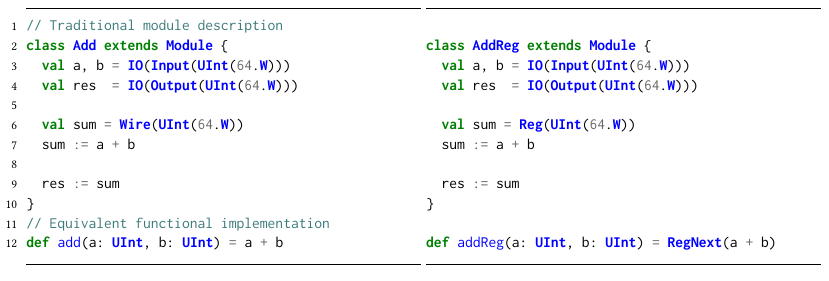}

    \begin{minipage}[t]{0.48\textwidth}
      \vspace*{-1.5em}
        \captionof{listing}{Combinatorial addition in vanilla Chisel}
        \label{lst:chisel:add}
    \end{minipage}
    \begin{minipage}[t]{0.46\textwidth}
      \vspace*{-1.5em}
      \captionof{listing}{Registered addition in vanilla Chisel}
      \label{lst:chisel:addreg}
    \end{minipage}
  \end{center}
\end{figure}

To introduce the need for architectural parameterization, we consider an adder module written in vanilla \emph{Chisel}~\cite{bachrach2012chisel}, a \emph{Scala}-embedded Hardware Construction Language (HCL).
Its usage in the following examples is very close to traditional HDLs, with similar keywords: \chisel{Module}, \chisel{Input}, \chisel{Output}, \chisel{Reg}, \chisel{Wire}.
Listing~\ref{lst:chisel:add} describes a simple combinational adder.
Lines $2$--$10$ express it as a standard module.
Line $12$ gives the equivalent inlined function, as it will appear in the following examples for conciseness.

Let us assume that this first version of the adder meet the timing requirements of the current FPGA target, but struggles with another one.
To accommodate the second target, we insert a register breaking the combinational path from \chisel{a} and \chisel{b} inputs to \chisel{res} output.
Listing~\ref{lst:chisel:addreg} illustrates the creation of this minimalistic pipeline stage.
The most notable difference with traditional HDLs stands in the switch from the event-driven assignation paradigm to an object-oriented declaration of wires and registers.
Lines $6$ and $7$ of Listings~\ref{lst:chisel:add} and~\ref{lst:chisel:addreg} depict this behavior.
In both descriptions, the assignation remains the same (\chisel{:=}) at line $7$, but the declaration of signal \chisel{sum} at line $6$ defines the actual hardware primitive: either a register (\chisel{Reg}) or a wire (\chisel{Wire}). 

Line $12$ demonstrates the ability of \emph{Chisel} functions to generate actual hardware (such as wire and registers).
Here, the standard \chisel{RegNext} object is used to functionally insert a register to store the result of the combinational addition.
This is a substantial difference with traditional HDLs in which functions can only be used to describe parameter computations or purely combinational logic.

\begin{figure}[tbp]
  \begin{center}
    \includegraphics[width=0.45\textwidth]{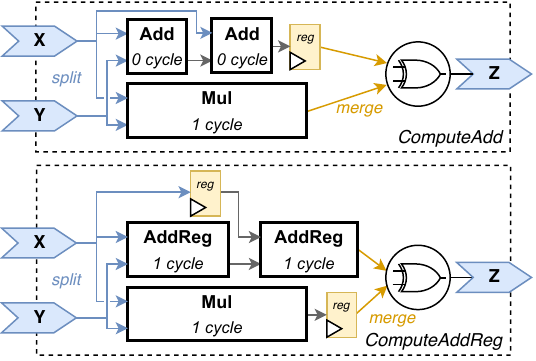}
    \caption{Potential integration issue with submodule latency evolution}
    \label{fig:integration}
  \end{center}
\end{figure}

\begin{figure}[thbp]
  \begin{center}
    \includegraphics[trim=43 452 45 92,clip, width=\textwidth]{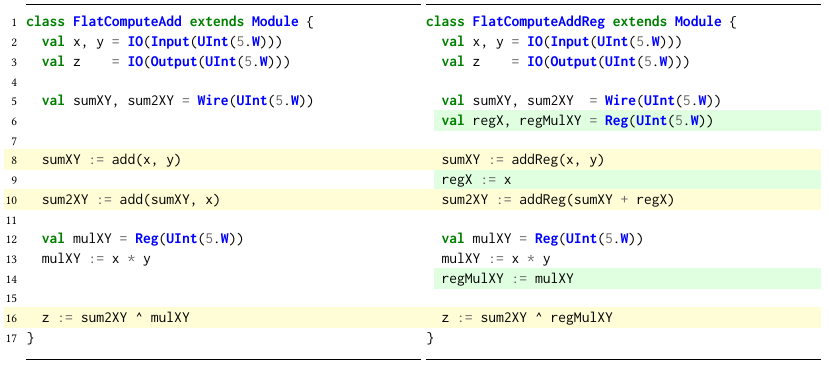}
    \begin{minipage}[t]{0.49\textwidth}
      \vspace*{-1.5em}
      \captionof{listing}{Flattened version of \emph{ComputeAdd} \\design, integrating non-registered adders}
      \label{lst:chisel:flatcomputeadd}
    \end{minipage}
    \begin{minipage}[t]{0.46\textwidth}
      \vspace*{-1.5em}
      \captionof{listing}{Flattened version of \emph{ComputeAddReg} \\design, integrating registered adders}
      \label{lst:chisel:flatcomputeaddreg}
    \end{minipage}
  \end{center}
  \vspace*{-1em}
\end{figure}

Thanks to \emph{Chisel} expressiveness, insertion of such a register requires updating a single keyword within the module body.
However, the most important side effect is revealed once this adder is integrated within a module hierarchy: the latency of the computation jumps from $0$ to $1$ clock cycle without further notice.
Figure~\ref{fig:integration} illustrates an integration example in which the expected result is $z = (2x + y) \oplus x y$.
Hardware implementation of $2x$ with an adder ($x+x$) rather than a shift ($x \ll 1$) is a deliberate choice in order to maintain the example conciseness while illustrating the critical issue of latency compensation.
The upper part of Figure~\ref{fig:integration} presents a first version of the design including two instances of the purely combinational \chisel{Add} module in parallel of a \emph{1-cycle} multiplier.
A flattened version of this design, based on functions rather than module instantiations, is described in Listing~\ref{lst:chisel:flatcomputeadd}.

The lower part of Figure~\ref{fig:integration} shows the update required to preserve the correctness of the overall design when using \chisel{addReg} instead of \chisel{add}.
Similarly, Listing~\ref{lst:chisel:flatcomputeaddreg} presents the corresponding flattened version, which inserts a register after each sum (1-cycle delay).
This latency increase is then compensated with the insertion of two additional registers: \chisel{regX} to delay \chisel{x} and \chisel{regMulXY} to delay \chisel{mulXY}.
Colored lines highlight differences from the previous \chisel{FlatComputeAdd} version: green background for additions and yellow background for modifications.
Among the $7$ lines of the original module body (lines $5$ to $16$ included) only $3$ remain unchanged whereas $3$ lines were added and $3$ were modified in the updated version, resulting in a $66\%$ rewrite of the original module body.

In practice the designer is left alone in charge of mentally inferring any latency changes as tools do not provide automated computation of this information.
These changes might require both local updates---as in these short code snippets---and global updates---throughout the hierarchy from the innermost modules to the top.
To preserve design correctness, latency changes must then be manually compensated and propagated still without any additional help from EDA tools.

The choice of inserting a register to break the combinational path might be driven by reusability needs, for example to accommodate the design to other FPGA targets.
While \emph{Chisel} environment would allow designers to parameterize the \chisel{Add} module to either insert a register or not, this change would still need be manually reflected to the surrounding components.
Not only the reusability need leads to architectural parameterization, but this parameterization requires some automation to become relevant at scale.
Moreover, in realistic use-cases, in which the latency is not always a few cycles but up to tens or hundreds of them, the choice of equivalent delay implementation becomes a typical architectural parameter.
In particular, implementations with shift-registers or memories can benefit from explicit target-dependent primitives to achieve the most efficient resource usage and best routability.

Another potential architectural parameter for pipelines is the definition of a handshake between stages, also referred as protocol signaling.
A given computation might need to be reused in various contexts, each requiring its own handshake system, such as the well-known protocols \emph{Ready-Valid}, \emph{Credit-Based}, or \emph{Carloni}~\cite{abbas2018latency}.
As the verbosity and complexity of the examples would highly worsen with protocol signaling usage, they are omitted here for conciseness.
Similarly to register insertion, introducing or swapping protocols on an existing hardware design is a laborious and error-prone task which requires some automation and pipeline modeling to be efficiently parameterized.

This minimalistic example showcases the intricate link between computation and implementation in traditional descriptions, and the global impact caused by even the smallest local modification.
Two potential architectural parameters have been identified: hardware primitives for signal delay implementation and protocol signaling implementation.
However, exposing these parameters requires first and foremost a proper modeling of the pipeline, in order to provide the needed automation to reflect local architectural choice to the surrounding context.
To provide this parameterization and achieve the associated automation, we aim at modeling the pipeline descriptions around the two following concepts:
\begin{description}
  \item[Latency-awareness] Modeling signal synchronization requirements to automate the insertion of configurable architectural elements
  \item[Protocol-polymorphism] Protocol-agnostic descriptions with elaboration-time configuration
\end{description}

\subsection{Relations as First-Class Objects}
Both latency and protocol express the \emph{relations} between various parts of a design in terms of synchronization.
To define the boundaries of such parts, we consider groups of coherent signals, i.e. signals on which direct combinational operations would yield meaningful results.
Such signals are said to be synchronized and we call each such part of the design a \emph{TimeZone}.
These TimeZones and their respective relations are the fundamental elements of our custom pipeline synchronization model.

Several models of pipelines synchronization have previously been detailed in the literature~\cite{leiserson1991retiming,sakallah1993synchronization} and compare to our model.
In particular, our model is related to Leiserson's~\cite{leiserson1991retiming} graph approach for retiming. 
We also introduce a graph representation of the pipeline as a support to illustrate how data signals relate to each other.
The original study leverages this graph representation to model combinational latencies between registers and aims at minimizing clock periods by moving some registers to cut the longest combinational paths.
We instead model the relations (latency and protocol) between groups of coherent signals and aim at making configurable the synchronization paths between these so-called TimeZones.
Our approach differs both in the meaning of the nodes (combinational operator vs a group of signals) and our usage of the graph.
In the original approach, the graph is fully constrained from the description, and the manipulations aim at maintaining its overall consistency while updating the value of edges.
In our approach, the initial version of the graph, as it is described by the designer, is containing some partially defined edges that we aim at automatically implement.

As we do not intend to provide automated scheduling based on signal dependencies, the main relations between successive TimeZones are required to be properly defined.
However, we claim that relations between TimeZones do not \emph{always} need to be exhaustively described.
In particular, the implementation of the trivial -- but usually verbose -- signal propagations along pipeline operations can be automated and made configurable.
Similarly, protocol signaling implementation can be automated, based on the synchronization model.
To establish a signal synchronization model able to provide these properties, we retain the following definitions:

\begin{description}

\item[Signal \emph{(electrical)}]
Carrier of electrical variations which encode some pieces of information.\\
Within a circuit, a signal is known by its \emph{unique} name, and its sole property is its \emph{type}, ranging from simple boolean to complex structured types.
A signal has a single driver (source) and might drive one or multiple components of the circuit (sinks).

\item[TimeZone]
Group of fully synchronized signals.\\
Direct combinational operations on these signals are sensible from a functionality perspective, without requiring any additional delay or handshakes.
In particular, all those signals share the exact same protocol signals, e.g. validity and back-pressure signals.

\item[Relation]
Synchronization requirements between two given TimeZones.\\ 
A relation does not model the actual operational logic actually connecting the signals.
Instead, it carries two properties: latency and protocol signaling.
Undefined protocol signaling means an absence of underlying requirement and hence the ability to configure any protocol for actual implementation.
Undefined latency means a \emph{missing relation} that needs to be solved by the synchronization resolution algorithms, based on user-configured implementation strategies.
In the current model, such \emph{missing relations} are only allowed to express propagations of a signal between its original driver and delayed downstream references (see Signal \emph{(pipeline)}).
A relation does not always model an actual hardware connection between signals. 
In particular, abstract \emph{equivalent Relations} can be composed of successive relations.

\item[PipeStep]
Oriented functional transition between a source TimeZone and a sink TimeZone.\\
It defines both the relation between the two TimeZones and the actual hardware logic connecting signals of each TimeZone.
A PipeStep can use any signal available in its source TimeZone and might declare new signals in its sink TimeZone.
A PipeStep generalizes the well-defined concept of pipeline stage which consists of a combinational computation and its associated pipelining register.
A PipeStep might describe a usual pipeline stage, but it can also describe a purely combinational assignation, or multi-cycle computations (typically to model a memory access).

\item[Pipeline]
Explicitly scheduled description of hardware operations as successive PipeSteps.\\
A pipeline consists of a collection of PipeSteps and their associated TimeZones, with sink TimeZones of PipeSteps being the source TimeZones of downstream PipeSteps.
A pipeline may consist of multiple branches as follows.
Multiple PipeSteps sharing a same source TimeZone model a \emph{split} of the pipeline between distinct \emph{branches}.
Conversely, multiple PipeSteps sharing a same sink TimeZone model a \emph{merge} of pipeline \emph{branches}.

\item[Signal \emph{(pipeline)}]
Named element of a pipeline.\\
Refers both to the original \emph{(electrical)} signal associated to a source TimeZone and all its delayed versions, used in downstream PipeSteps and TimeZones.
Within a pipeline, a signal describes multiple electrical signals, each associated to a local TimeZone. 
The delay between the respective sequences of values of each electrical signal match the relations between their respective TimeZones.
In the current model, \emph{missing relations} can only describe equivocal downstream references to a previously defined source signal.
In particular, all new signal must be created as part of a TimeZone with explicit relations to the surrounding TimeZones.

\item[Synchronization Model]
Directed Acyclic Graph (DAG) composed of TimeZones as nodes and relations as edges.
A first partial graph---i.e. containing \emph{missing relations}---is obtained from the pipeline description.
It serves as intermediate representation of the hardware to obtain a fully-synchronized and implemented circuit thanks to synchronization resolution algorithms presented below.

\end{description}

\subsection{Circuit Resolution Process}

Based on the previous definitions, synchronization of the circuit is obtained through the following 3-stage process:
\begin{description}
  \item[1. Model Construction] From an \emph{ad-hoc} pipeline-oriented description of the circuit, this first step produces the abovementioned synchronization model.
  Model construction is further detailed in Section~\ref{sec:model:construction}.
  \item[2. Synchronization Resolution] Iterating on the synchronization model as an intermediate representation, this second step computes all the additional hardware and connections required to achieve circuit synchronization.
  Various resolution strategies and their implementations are described in Section~\ref{sec:sync:reso}.
  \item[3. Circuit Update] Based on the results of synchronization, this final step consists in producing the actual exhaustive circuit description.
\end{description}

The ability to draw a clear interface between these steps highlights the complete independence of the synchronization model and its associated resolution algorithms from the concrete description of the circuit.
In particular, although this paper presents an implementation of this process based on \emph{Chisel} in the next section, all the formalism and resolution strategies are independent of the implementation language.

\section{Framework Implementation}
\label{sec:framework:implementation}

As input of the synchronization model, the initial pipeline description aims at retaining only the minimal amount of implementation details required to ensure a deterministic resolution of the synchronization.
In particular, to provide the desired architectural parameterization, previous sections described two kinds of implementation details that can be omitted:
\begin{itemize}
  \item Explicit propagation of signals from their definition TimeZone to their usage,
  \item Handshake mechanisms and protocol signals.
\end{itemize}

The less implementation details are hardcoded in the pipeline description, the more flexible and configurable it can become with the proper design methodology.

\subsection{Pipeline Design Methodology}

\begin{figure}[tbp]
  \begin{center}
    \includegraphics[width=0.38\textwidth]{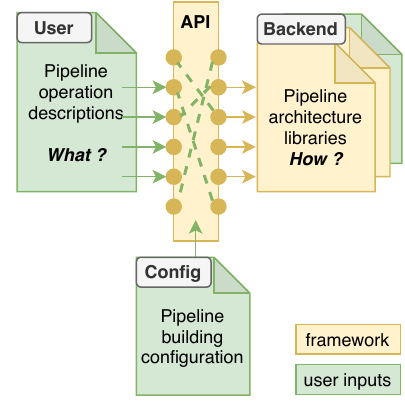}
    \caption{Decoupled Pipeline Description Methodology}
    \label{fig:api}
  \end{center}
\end{figure}

Critical implementation choices are often dictated by the applicative context and the available hardware target.
This section describes a design methodology which decouples the pipeline description from its architectural configuration.
Figure~\ref{fig:api} illustrates our approach, built around the synchronization model defined in the previous section, as a triple-sided Application Programming Interface (API):
\begin{description}
  \item[User API] Pipeline structure and operations are user-described as a dedicated set of classes and objects.
  This API defines the syntax and the primitives which creates the synchronization model.
  \item[Backend API] Synchronization resolution algorithms and protocol implementation operate on the synchronization model to obtain a consistent circuit.
  This API contains the algorithms described in Section~\ref{sec:sync:reso} and users retain the ability to extend them or develop their own ones as additional libraries,
  \item[Config API] An elaboration-time \emph{configuration API} provides fine-grained configuration abilities in the mapping between \emph{user-facing API} and \emph{backend API}.
  It drives the actual code generation from classes and objects available in user and backend APIs.
\end{description}

The signal synchronization model is the common interface between user-API---describing signal interactions---and backend-API---in charge of guaranteeing their synchronizations.
Section~\ref{sec:sync:reso} details the algorithms used to provide signal synchronization on the back-end side.
The remaining part of the current section first focuses on the concrete implementation of the \emph{User API} and then gives some practical example of the \emph{Config API}.

\subsection{User API: Model Construction}
\label{sec:model:construction}

The actual implementation of the \emph{User API} is a Scala library defining pipeline-oriented objects and properties, intended to be used within a \emph{Chisel} hardware generation context.

\begin{description}

\item[Pipe]
Created as an explicit object \chisel{Pipe(input, name)}, a pipeline is a container for pipeline branches.
It defines a default main branch, with an initial TimeZone containing the signals available from \chisel{input} relation.
A pipeline behaves as a branch, but also defines a \chisel{build} method which triggers the actual elaboration of the synchronization model from the description.

\item[Branches]
Succession of PipeSteps, branches most notably provide the \chisel{split} and \chisel{merge} methods.

\item[Steps]
Defined as anonymous functions \chisel{(p,n) => {}}, PipeSteps defines the relation and the hardware connections between a previous TimeZone \chisel{p} and a new one \chisel{n}.

\item[Signal Ubiquity Property]
In the synchronization model, a signal is a name referring both to a source signal and all downstream use of this signal, with appropriate latency compensation.
In the pipeline description, any downstream reference by name to a previously defined signal is to be provided to the designer as an appropriately delayed version of the original source, without requiring explicit descriptions of delays.
To guarantee this property, the pipeline is expressed in a Single-Static-Assignment (SSA) form.
Each signal is defined and connected only once in a given TimeZone, and can then be used multiple times downstream in the pipeline.
As a pure SSA form would conflict with the generation needs (loops, functions) of advanced circuit design, SSA form is enforced within contained areas.
These areas can then be connected to one another with explicit mapping of signals by name, similarly to module instantiation patterns found in traditional hardware descriptions.

\end{description}

\begin{figure}[htbp]
  \begin{center}
    \includegraphics[trim=43 347 45 84,clip, width=\textwidth]{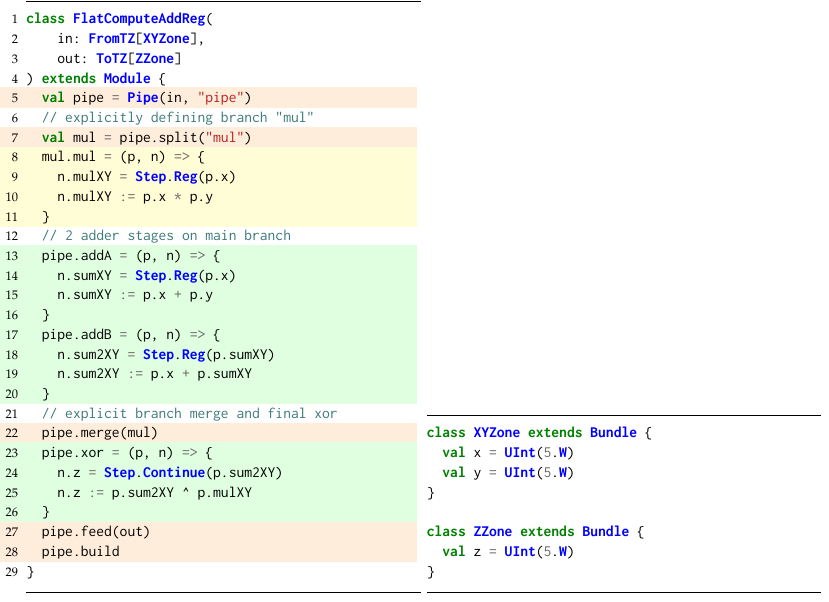}

    \begin{minipage}[t]{0.49\textwidth}
      \vspace*{-1.5em}
      \captionof{listing}{Running example implemented with our pipeline-oriented framework based on Chisel HCL}
      \label{lst:code:running:chisel}
    \end{minipage}
    \begin{minipage}[t]{0.46\textwidth}
      \vspace*{-1.5em}
      \captionof{listing}{I/O TimeZones}
      \label{lst:chisel:TimeZone:ios}
    \end{minipage}
  \end{center}
\end{figure}

The application of this design pattern to the motivating example of Section~\ref{sec:architectural:param} is presented in Listing~\ref{lst:code:running:chisel}.
To facilitate comprehension, code background is colored with the following correspondences:
\begin{enumerate*}
  \item orange for explicit operation on the pipeline object and its branches (\chisel{split}, \chisel{merge}, \chisel{build}),
  \item green for operations (PipeSteps) on the main (default) branch of the pipeline, and
  \item yellow for PipeSteps on the explicit \chisel{mul} branch.
\end{enumerate*}

This pipeline-oriented version is equivalent to the vanilla Chisel version \chisel{FlatComputeAddReg} depicted in Listing~\ref{lst:chisel:flatcomputeaddreg}, which inserts a register after each adder.
On such a small example, the number of lines of code (LoCs) considerably increases as the concrete definition of the pipeline, its branches and the steps each introduce a few lines of overhead.
As the actual operational content of the pipe is quite sparse (only 4 combinational operators), all this overhead seems very excessive for the task.
However, substantial LoC reduction can be observed in more advanced examples, such as the one presented in Section~\ref{sec:xp:indus}.  
This initial example mainly aims at illustrating the main concepts of our approach:
\begin{itemize}
  \item Pipeline and branching operations are fully explicit and steps are quickly identified,
  \item Relative scheduling of each new TimeZone is explicitly described by the \chisel{Step} object, either as \chisel{Wire} or \chisel{Reg},
  \item Signal propagation and handshakes are omitted,
  \item Any local change of such user-described relation between TimeZone is expected to maintain the proper synchronization of all propagated signals.
\end{itemize}
This last property is guaranteed by the synchronization algorithms detailed in the next section.
The current subsection further details the syntax of the \emph{User API} and is followed by several generation examples based on this single description.

Listing~\ref{lst:code:running:chisel} describes practical usage of the main objects and classes provided by the user API.
The concrete class \chisel{Pipe} is the core of this API.
Its instantiation at line $5$ specifies the source relation (\chisel{in}) of the pipeline, with type \chisel{XYZone} whose declaration in Listing~\ref{lst:chisel:TimeZone:ios} provides signals \chisel{x} and \chisel{y}.
These two input signals are natively available to all downstream operations, as seen on lines $10$, $15$ and $19$. 

The \chisel{Pipe} object provides a default (main) branch on which operations are successively recorded as explicit functions (PipeSteps).
To implement the example, three of such PipeSteps are declared on the main branch, on lines $13$--$16$ (first adder), $17$--$20$ (second adder) and $23$--$26$ (final xor).
Each PipeStep consists of an anonymous function taking two TimeZones as arguments, here \chisel{p} for previous (source) and \chisel{n} for next (sink).
In the function body at line $14$, the \chisel{Step.Reg} function describes the creation of a new signal \chisel{sumXY} in TimeZone \chisel{n}, with a relation from TimeZone \chisel{p} of $1$-cycle latency.
While this syntax is very similar to a usual register description, it also defines the relation between TimeZones and their signals in the synchronization model.
Actual connection of this signal, as a result of the addition of \chisel{x} and \chisel{y} signals, is described at line $15$.
Similarly, at line $24$, in the function body of the final xor operation, the \chisel{Step.Wire} function describes the creation of a new signal \chisel{z} in TimeZone \chisel{n}, with a relation from TimeZone \chisel{p} of $0$-cycle latency, equivalent to a usual wire---i.e. combinational---stage.
During pipeline elaboration, TimeZones are created sequentially by executing the PipeSteps one after another.
The newest TimeZone serves as previous TimeZone for the following PipeStep which itself creates another TimeZone.

The \chisel{Pipe} object also provides explicit pipeline-oriented operations such as creation of new branches, split from the main branch at its current state, i.e. after the last PipeStep.
The split creating \chisel{mul} branch occurs on line $7$, at the very beginning of the pipeline and creates a fully parallel branch.
Lines $8$--$11$ define a PipeSteps on this new branch.
It is then merged into the main branch on line $22$.
A last operation, for which signals from both branches are now available, occurs on lines $23$--$26$.

To forward the computation result, on line $27$, the \chisel{Pipe} instance declares connection to external TimeZone at its current state: here, signal \chisel{n.z} declared in last PipeStep \chisel{pipe.xor} is connected to sink relation \chisel{out}.
Underlying type of this relation \chisel{ZZone} only contains the signal \chisel{z} as described in Listing~\ref{lst:chisel:TimeZone:ios}.

Finally, line $28$ triggers the pipeline elaboration and actual build of the synchronization model, based on all the operations previously recorded from initialization to operations on respective branches to final feed.
Actual TimeZones objects are created and passed as arguments of the PipeStep functions to record signal creation and usage within the synchronization model.

\begin{figure}[htbp]
  \vspace*{-10pt}
  \centerline{
    \begin{subfigure}{0.70\textwidth}
    \includegraphics[width=\textwidth]{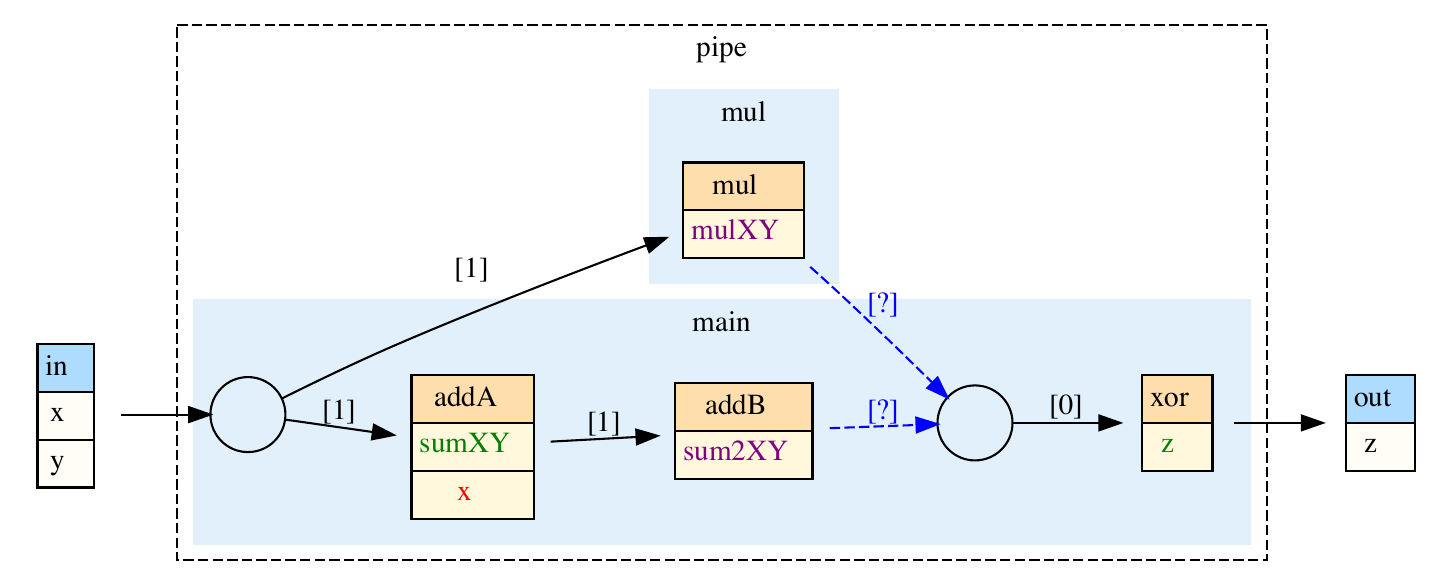}
    \caption{Graph representation corresponding to Listing~\ref{lst:code:running:chisel}}
    \label{fig:reso:intent}
  \end{subfigure}
  \begin{subfigure}{0.30\textwidth}
    \includegraphics[width=\textwidth]{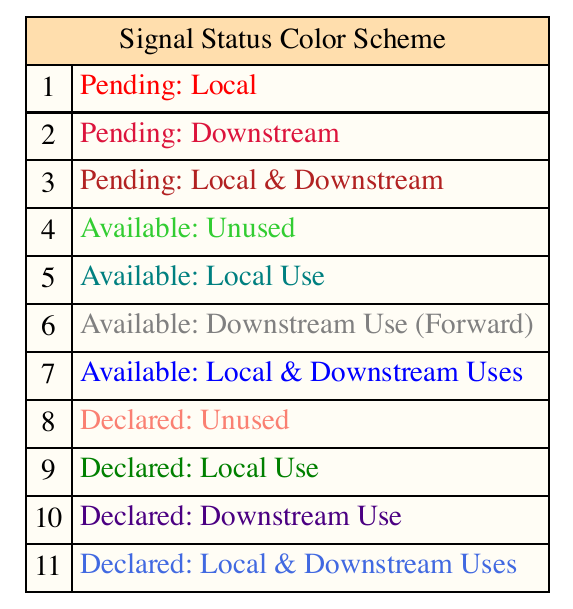}
    \caption{Signal Color Scheme}
    \label{fig:signal:legend}
  \end{subfigure}
  }
  \caption{Design Intent}
  \label{fig:reso:intent:all}
\end{figure}

At this stage, the pipeline description focuses exclusively on the design intent: signal declarations, connections and operations between these signals.
Some register stages have been explicitly defined after each adder and multiplier, based on prior knowledge of target performance and application needs.
However, there are no mention of protocol signaling associated to any operation, and no register compensation, in particular to guarantee proper synchronization between \chisel{main} and \chisel{mul} branches.
Figure~\ref{fig:reso:intent} illustrates the corresponding design intent, with explicit \emph{split/merge} operations on two branches of the pipeline, while some relations are intentionally left unspecified.
Edge labels illustrate respective relation latencies, either unknown \emph{[?]} or explicitly defined here as registers \emph{[1]} or simple wires \emph{[0]}.
Each PipeStep results in a relation towards a node---its output \chisel{n} TimeZone---labelled with its name as table heading.
The node then lists all signals either defined by the upstream PipeStep or required by the downstream PipeStep.
A color code, presented in Figure~\ref{fig:signal:legend}, provides further insight on the current status of each signal in each TimeZone.
Here signal \chisel{sumXY}, which is defined by \chisel{addA} PipeStep and is immediately used by \chisel{addB} PipeStep, is colored in green corresponding to status $9$: \emph{Declared: Local use}.
Signal \chisel{x}, which is also used by \chisel{addB} PipeStep but is not provided by \chisel{addA} PipeStep, is colored in red corresponding to status $1$: \emph{Pending: Local}.

As a result, Figure~\ref{fig:reso:intent} illustrates the base synchronization needed to convert this simple design intent into a working circuit: delayed connection of signal \chisel{x} and balancing of branches \chisel{main} and \chisel{mul}.

\subsection{Config API: Model Parameterization}
\begin{figure}[ht]
    \begin{center}
    \vspace*{-0.5em}
    \includegraphics[trim=43 545 45 90,clip, width=\textwidth]{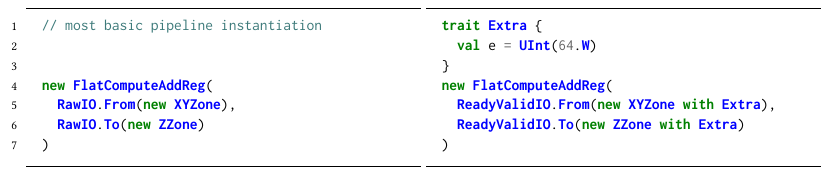}

    \begin{minipage}[t]{0.48\textwidth}
      \vspace*{-1.5em}
      \captionof{listing}{Framework generation without protocol}
      \label{lst:chisel:gen:raw}
    \end{minipage}
    \begin{minipage}[t]{0.46\textwidth}
      \vspace*{-1.5em}
      \captionof{listing}{Framework generation with \emph{ready-valid} protocol and extra signal}
      \label{lst:chisel:gen:readyvalid}
    \end{minipage}
  \end{center}
\end{figure}

The pipeline description presented in Listing~\ref{lst:code:running:chisel} is a hardware generator which requires parameterization to generate a circuit, here with \chisel{in} and \chisel{out} TimeZones.
This elaboration-time parameterization enables designers to generate various circuits based on the same design intent and is leveraged here to provide two degrees of freedom:
\begin{enumerate*}
  \item protocol polymorphism, and
  \item extra signal propagation.
\end{enumerate*}
Protocol polymorphism is available as part of the relation definition while extra signal propagation is based on the ability to extend the surrounding \chisel{in} and \chisel{out} TimeZones for additional signal synchronization.
Listing~\ref{lst:chisel:gen:raw} illustrates the generation of the circuit in its minimal form, without protocol signaling (\chisel{RawIO}).

Generating a different circuit based on ready/valid handshakes between each TimeZone only requires swapping \chisel{RawIO} for \chisel{ReadyValidIO}.
Listing~\ref{lst:chisel:gen:readyvalid} illustrates this protocol swap and also introduces the extra signal propagation by extending both \chisel{XYZone} and \chisel{ZZone} TimeZones with signal \chisel{e}.

\begin{figure}[htbp]
  \begin{center}
    \includegraphics[width=0.85\textwidth]{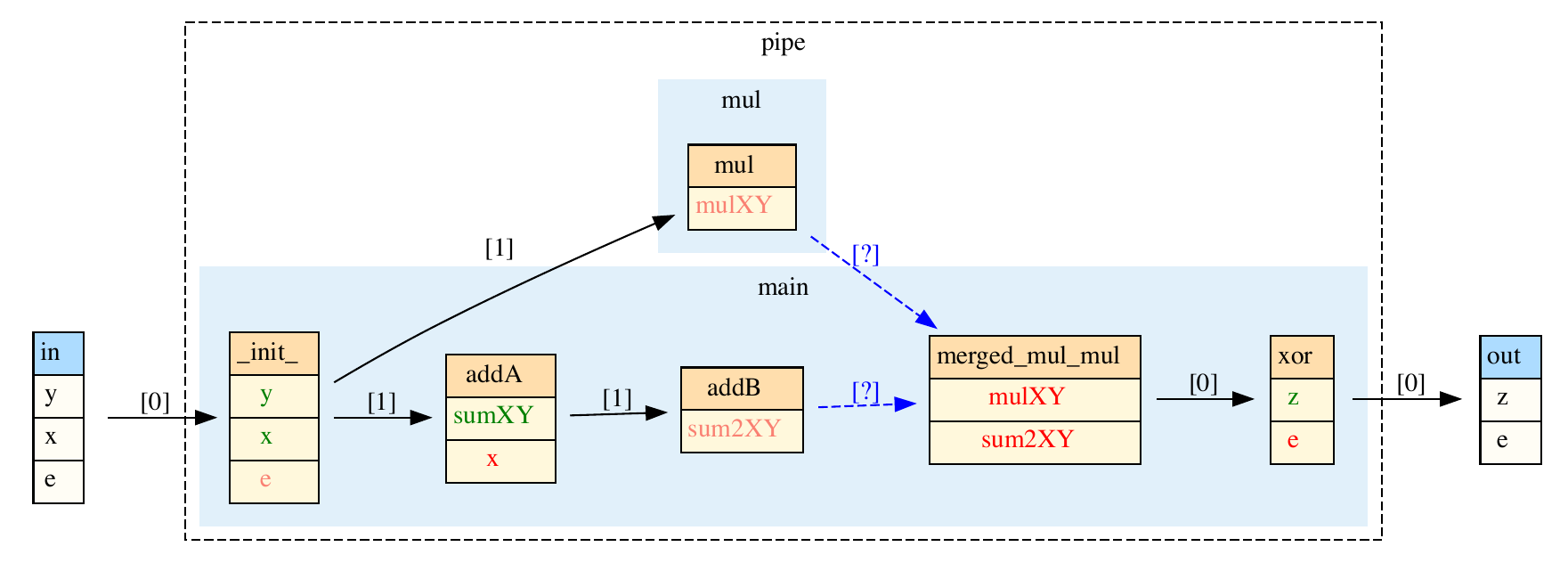}
    \caption{Generation Intent with Additional Signal \chiselBlack{e}}
    \label{fig:reso:nores}
  \end{center}
\end{figure}
Figure~\ref{fig:reso:nores} illustrates the corresponding generation intent, with the introduction of signal \chisel{e}.
Protocol signals, namely \emph{ready} and \emph{valid}, are automatically inserted at both input and output and are intentionally omitted on the graph representation as they are already implicitly part of the relations and TimeZones.

Both changes are very concise, yet yielding very different circuits without requiring any modification of the original pipeline description.

Similarly, the pipeline description itself is resilient to local changes without requiring remote manual compensations as previously illustrated with the register insertion example in Section~\ref{sec:dec:context}.
To toggle register stage insertion off, the \chisel{Step.Reg} function, used to define values \chisel{sumXY} and \chisel{sum2XY}, respectively on lines $14$ and $18$, can be replaced by the \chisel{Step.Wire} function.
The version with \chisel{Step.Reg} and \chisel{RawIO} is equivalent to the Chisel module \chisel{FlatComputeAddReg}, while the \chisel{Step.Wire} version is equivalent to the Chisel module \chisel{FlatComputeAdd}.
These functions define appropriate relations between signals, allowing automatic latency propagation, path balancing and latency-awareness for surrounding descriptions.
Instead of in-depth refactoring for each use case, the module is now configurable by protocol and pipeline stages can be converted from register to wire with very little and local-only effort.

The algorithms leveraged to provide these advanced generation features, based on the resolution of this elaborated synchronization model, are detailed in the next section.

\section{Model Resolution}
\label{sec:sync:reso}

\subsection{Base Principles}
\begin{figure}[tbp]

  \hspace*{0.90em}
  \centerline{
    \begin{subfigure}{0.55\textwidth}
      \includegraphics[width=1.05\textwidth,center]{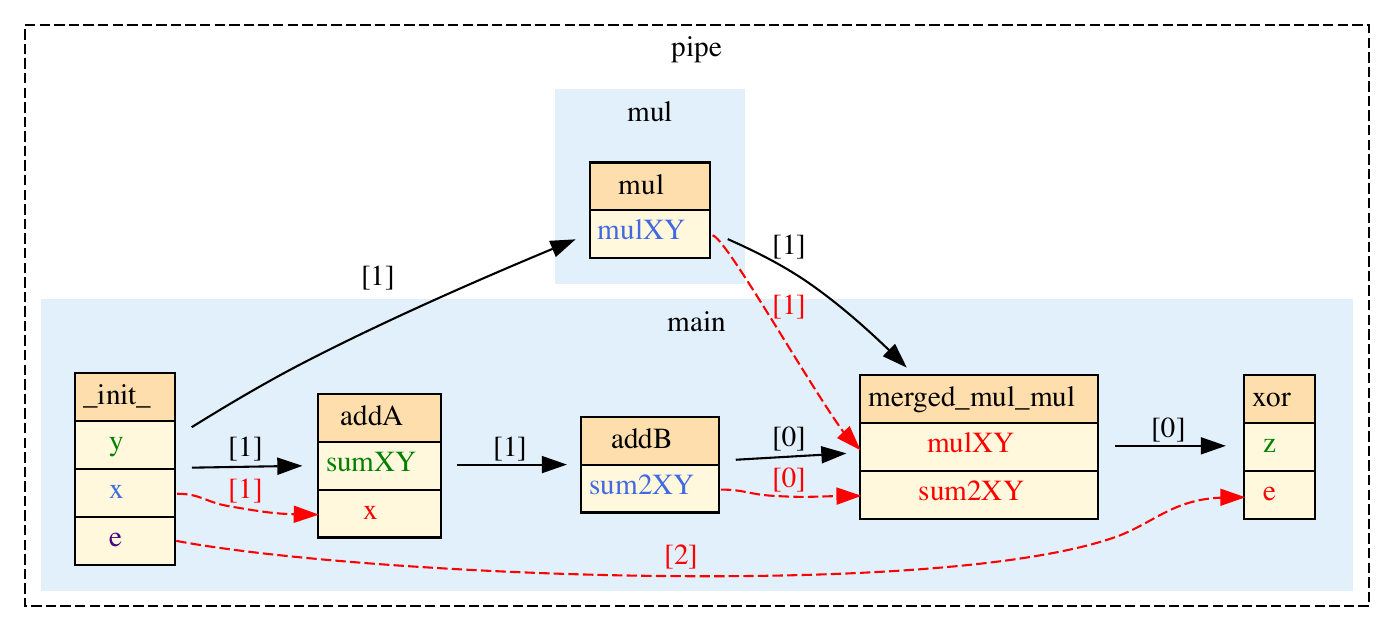}
      \captionsetup{justification=centering}
      \caption{Merge resolution without signal propagation\\ \footnotesize \emph{red dash: missing relations}}
      \label{fig:reso:noprop}
    \end{subfigure}
    \begin{subfigure}{0.55\textwidth}
      \includegraphics[width=0.95\textwidth,center]{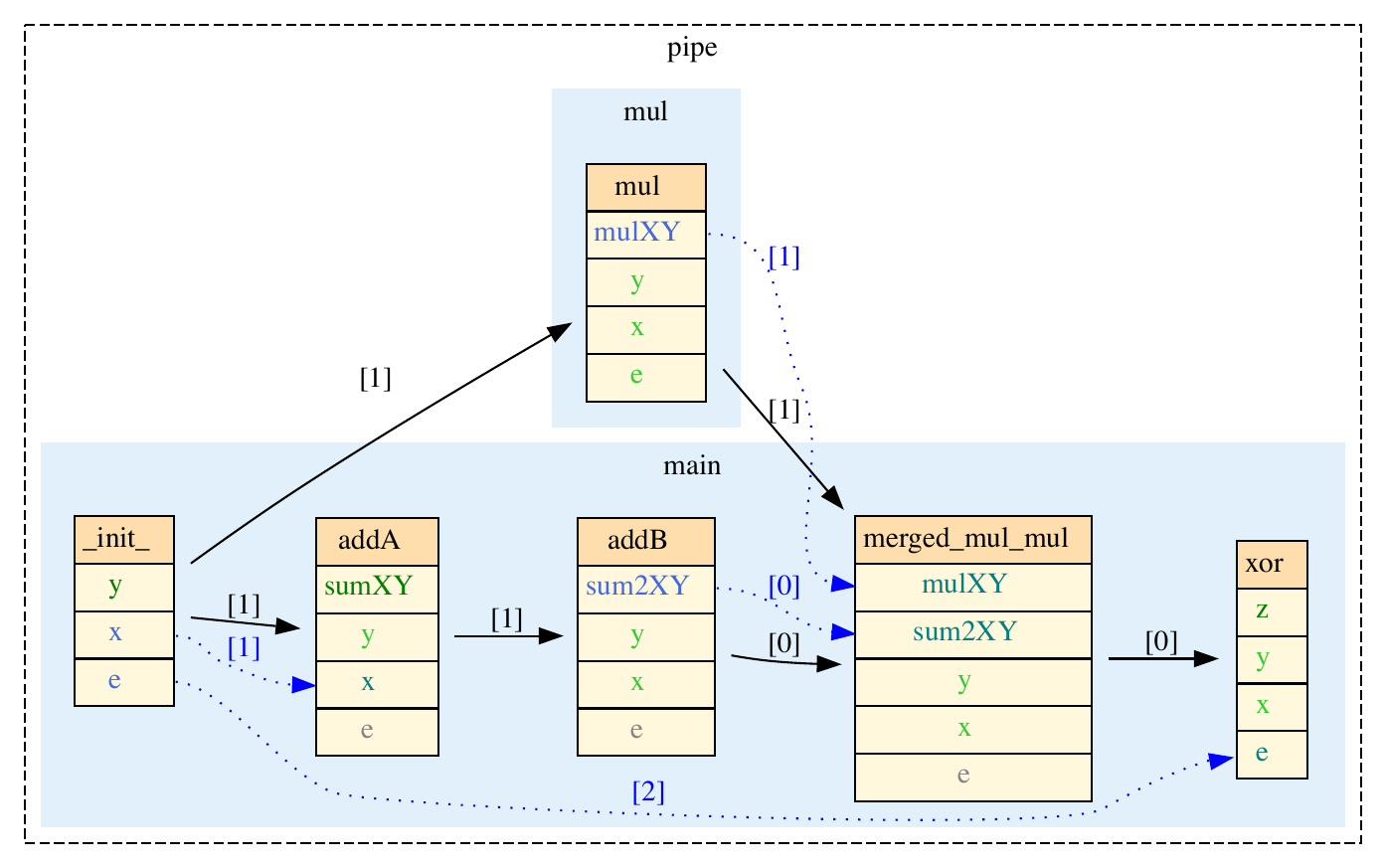}
      \caption{Exhaustive Forward Propagation Strategy}
      \label{fig:reso:exforward}
    \end{subfigure}

  }

  \vspace*{2em}

  \hspace*{0.50em}
  \centerline{
  \begin{subfigure}{0.55\textwidth}
    \includegraphics[width=0.95\textwidth,center]{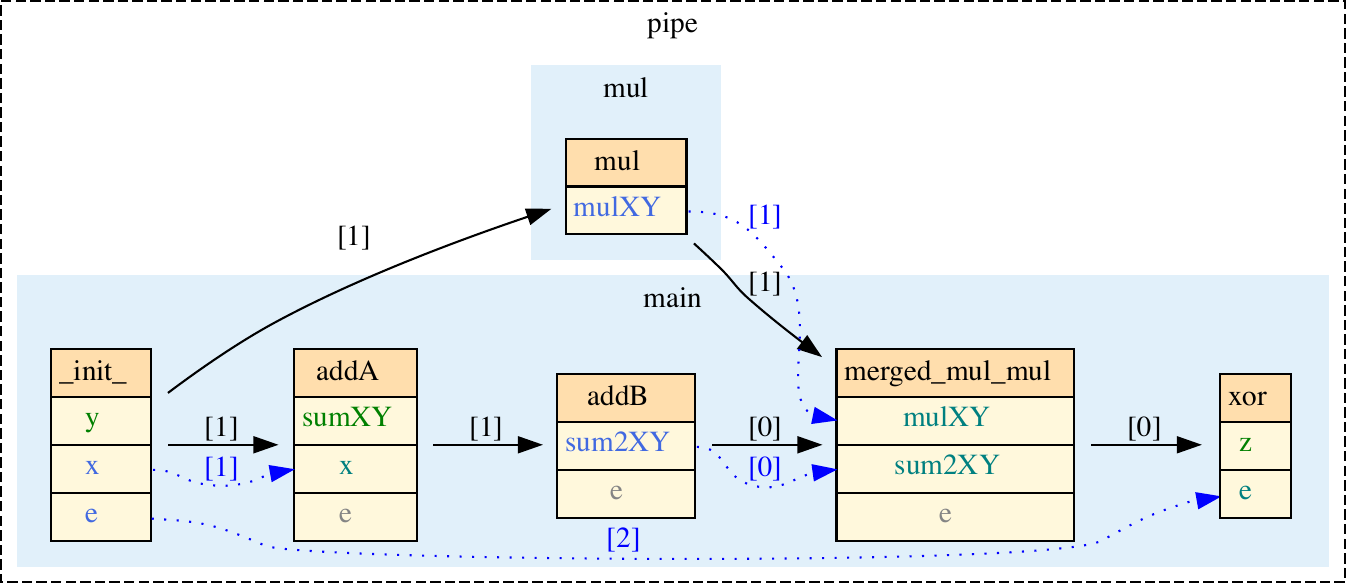}
    \captionsetup{justification=centering}
    \caption{Peer-to-Peer Propagation Strategy \\ \footnotesize \emph{blue dot: transitive implementation}}
    \label{fig:reso:p2p}
  \end{subfigure}
  \begin{subfigure}{0.55\textwidth}
    \includegraphics[width=0.95\textwidth,center]{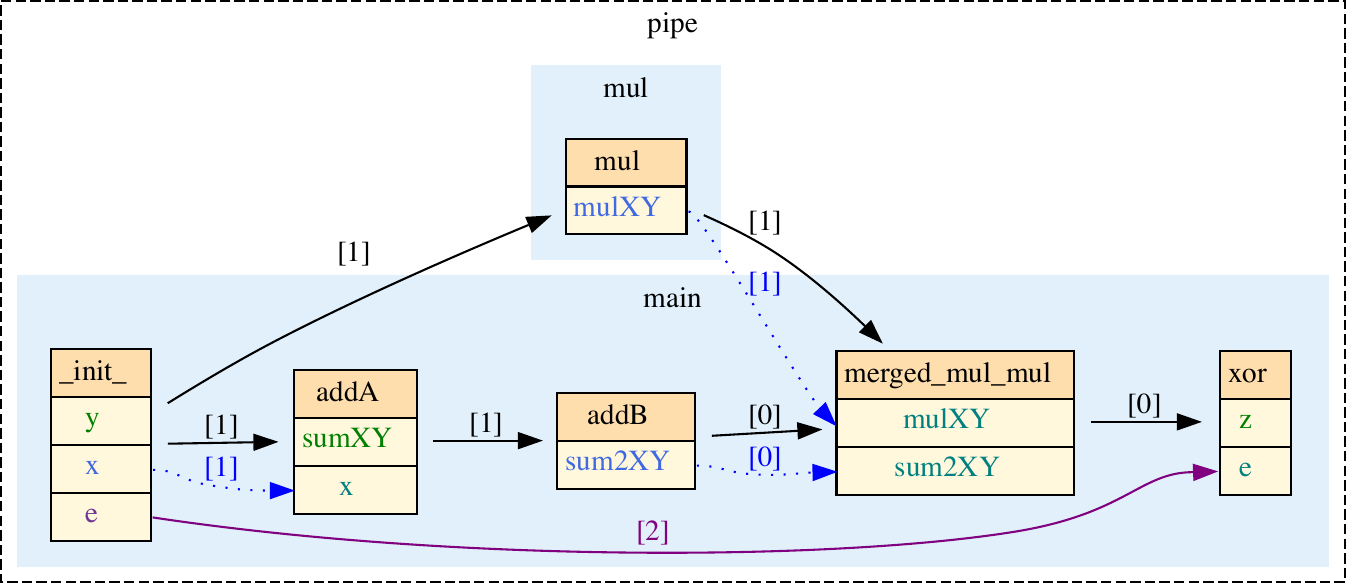}
    \captionsetup{justification=centering}
    \caption{Direct Propagation Strategy\\ \footnotesize \emph{plain purple: direct implementation}}
    \label{fig:reso:direct}
    \vspace*{0.5em}
  \end{subfigure}
  }
  \caption{Resolution steps and options for the running example}
  \label{fig:reso:all}
\end{figure}

A successful model resolution consists in producing a balanced graph without any relation left unspecified.
The synchronization model obtained after pipeline elaboration, depicted in Figure~\ref{fig:reso:nores}, serves as input to resolution algorithms.
It contains intentional omissions aimed at providing parameterization, meaning that the resolution algorithms will need to take some parameters into account during the resolution process.
The very first parameter is the algorithm itself, as the signal propagation can be achieved in multiple ways.
The following subsections each detail such solutions. 
We nonetheless retain a common overall architecture, as follows:
\begin{enumerate}
  \item A first set of analysis compute the equivalent latencies for each path and balance branches at their merge point, fully characterizing all missing relations (Figure~\ref{fig:reso:noprop}),
  \item The resolution algorithms creates latency-equivalent hardware in place of equivalent relations as illustrated in Figures~\ref{fig:reso:exforward}, \ref{fig:reso:p2p} and~\ref{fig:reso:direct},
  \item Then, a transformation is in charge of propagating and connecting protocol signals uniformly across the balanced graph,
  \item A final pass validates overall synchronization consistency, providing early feedback to the designer.
\end{enumerate}


\paragraph{Pre-requisite}
The algorithms presented below assume the availability of some base operations on the synchronization model graph.
This graph is expected to be a Directed Acyclic Graph (DAG), i.e. edges (relation) are oriented and there are no cycles in the graph.
Such cycles would have no meaning in our applicative context, as they would defeat the ability to provide throughput guarantees.
Nodes can hence be ordered, and the synchronization model can provide a forward path among nodes (TimeZone) from a unique input to one or several outputs.
Similarly, this order can be reversed to provide a meaningful backward path from outputs to input.
In practice, the forward path is created with the following traversal: from input TimeZone on main branch, proceed with the next TimeZone until a split, in which case proceed with external branches first, then continue on the initial main branch.
For our running example forward order is the following:
\begin{enumerate*}
  \item \emph{\_init\_},
  \item \emph{mul} (on branch \emph{mul}),
  \item \emph{addA},
  \item \emph{addB},
  \item \emph{merged\_mul\_mul},
  \item \emph{xor}.
\end{enumerate*}
The backward path order is simply the reversed version of this list.
Such traversal strategies ensure that all TimeZones are processed in a consistent order in terms of signal declaration and usage: the forward order ensures signal declarations to be processed before any usage whereas backward order ensures all usage to be inspected prior to the declaration, which is convenient for single-pass signal propagation algorithms.

Graph edges (relations) can be composed to form equivalent relations.
Each signal within each TimeZone retains an availability property, representing its usage and connection, as illustrated with corresponding colors on Figure~\ref{fig:reso:intent:all}.
With these two properties, \emph{equivalent missing relations} can be defined from a TimeZone where a signal is available to a TimeZone where it is used.
For example, missing relation for signal $e$ depicted in Figure~\ref{fig:reso:noprop} is based on the composition of successive edges from \emph{\_init\_} to \emph{xor}.
More precisely, Figure~\ref{fig:reso:noprop} shows the graph representation, still unresolved and ready for use by the following signal propagation algorithms.

\subsection{Resolution Algorithms}
This section does not only aim at demonstrating the ability to resolve the synchronization, but it also introduces the ability to create any custom resolution algorithm, finely tailored to application needs.
To that extend, the following paragraphs present the definition of a simple yet efficient single-pass resolution algorithm through 3 successive iterations.
Due to its additional parameterization capabilities, the final algorithm is selected for the industrial experimentation detailed in Section~\ref{sec:xp:indus} and~\ref{sec:xp:fifo:explo}.
For the sake of conciseness, the pseudocode illustrating the resolution strategies take branches as inputs instead of the full graph, hence not detailing split and merge management.

\paragraph{Exhaustive Forward Signal Propagation}
The first naive approach consists of forwarding all declared and used signals from TimeZone to TimeZone, indiscriminately of actual downstream usage.
Algorithm~\ref{algo:naive} details its logic.
The number of signals forwarded along the pipeline grows linearly stage after stage and a lot of unused hardware is generated.
Figure~\ref{fig:reso:exforward} illustrates this phenomenon, resulting in many available but unused signals across TimeZones.
Synthesis tools might be able to partially remove this \emph{dead-code} in simple cases, such as unconnected wires and registers, however useless signals won't necessarily be stripped out if they are stored in a memory, e.g. in a FIFO explicitly implemented with a vendor primitive.
While demonstrating a very simple way to deterministically solve the synchronization model, this naive approach exhibits poor results and must be improved to eradicate unnecessary hardware overhead.

\begin{algorithm}[ht]
  \small
  \DontPrintSemicolon
  \SetKwFunction{addAndConnect}{addAndConnect}
  \SetKwInOut{Input}{input}\SetKwInOut{Output}{output}
  \SetKw{in}{in}
  \SetKw{not}{not}

  $DeclaredSignals := []$\;
     \ForEach{$TimeZone$ \in $Branch$}
     {
        \ForEach{$Signal$ \in $TimeZone$}{
          \If{$Signal$ \not \in $DeclaredSignals$}{
            $DeclaredSignals \mathrel{+}= Signal$\;
          }
        }
        \ForEach{$Signal$ \in $DeclaredSignals$}{
          \If{$Signal$ \not \in $TimeZone$}{
            $TimeZone$.\addAndConnect{$Signal$}\;
          }
        }
     }

  \caption{Naive approach: exhaustive forward}
  \label{algo:naive}
\end{algorithm}

\begin{algorithm}[ht]
  \small
  \DontPrintSemicolon

  \SetKwFunction{createConnectionTo}{createConnectionTo}
  \SetKwFunction{createPlaceHolder}{createPlaceHolder}
  \SetKwFunction{updateTarget}{updateTarget}
  \SetKwFunction{isAvailable}{isAvailable}
  \SetKwInOut{Input}{input}\SetKwInOut{Output}{output}
  \SetKw{in}{in}
  \SetKw{reversed}{reversed}
  \SetKw{not}{not}

  $MissingSignals := []$\;
  \ForEach{$TimeZone$ \in \reversed $Branch$}{
    \ForEach{$Signal$ \in $MissingSignals$}{
      \If{$Signal$ \not \in $TimeZone$}{
        $TimeZone$.\createPlaceHolder{$Signal$}
      }
      $TimeZone$.\createConnectionTo{$Signal$}

      \If{$Signal$.\isAvailable{$TimeZone$}}{
        $MissingSignals \mathrel{-}= Signal$\;
      }
      \Else{
        $MissingSignals.\updateTarget{Signal}$\label{algo:backward:peer2peer:updateTarget}
      }
    }
    \ForEach{$Signal$ \in $TimeZone$}{
      \If{\not $Signal$.\isAvailable}{
        $MissingSignals \mathrel{+}= Signal$
      }
    }
  }

  \caption{Backward Peer-to-Peer approach:\newline \emph{Requesting unavailable signals to previous TimeZone}}
  \label{algo:backward:peer2peer}
\end{algorithm}

\paragraph{Peer-to-Peer Backward Propagation}
To avoid propagating unwanted signals, the algorithm must be aware of in which stage each signal is expected.
Backward graph traversal provides the ability to list signal usage before their declaration.
Taking advantage of this principle, algorithm~\ref{algo:backward:peer2peer} details such a reversed resolution, with a peer-to-peer approach somehow similar to the first algorithm.
However, in this backward \emph{peer-to-peer} approach, unavailable signals are requested if needed to upstream TimeZones rather than being automatically pushed by them.
For conciseness in the algorithm presentation, $Signal$ objects are to be understood as by-name references to fully-qualified hardware signals associated to their respective TimeZones.
If the current TimeZone cannot provide some pending requested signals, requests are forwarded to the closest upstream TimeZone, i.e. the next to be processed by the backward traversal of the graph.
In practice, a placeholder signal is created within the current TimeZone, this placeholder is connected to the signal placeholder from the downstream requesting TimeZone, and finally the request is updated with actual reference to the newly created placeholder.
This is the role of \emph{updateTarget} function on line~\ref{algo:backward:peer2peer:updateTarget}.
Figure~\ref{fig:reso:p2p} illustrates the resolution of the running example with this strategy.
Previously missing relations (red dashed arrows in Figure~\ref{fig:reso:noprop}) have been implemented through the existing relations between TimeZones and now appear as blue dotted arrows.
Despite valid, this approach lacks reflexivity, as the missing relations are never directly manipulated as objects, simply solved as a side effect of the backward propagation.
In particular, when a signal declaration is missing, the algorithm is only able to report the last TimeZone where it went missing, which is always the input TimeZone.
Moreover, in merge nodes, the algorithm has to request the signal to all upstream TimeZones, which can lead to ambiguous error reporting in case of concurrently defined signals.
An actual understanding of the precomputed missing relations is required to get rid of these limitations.

\begin{algorithm}[ht]
  \small
  \DontPrintSemicolon

  \SetKwFunction{createConnectionTo}{createConnectionTo}
  \SetKwFunction{createPlaceHolder}{createPlaceHolder}
  \SetKwFunction{updateTarget}{updateTarget}
  \SetKwFunction{isAvailable}{isAvailable}
  \SetKwInOut{Input}{input}\SetKwInOut{Output}{output}
  \SetKw{in}{in}
  \SetKw{reversed}{reversed}
  \SetKw{not}{not}

  $MissingSignals := []$\;
  \ForEach{$TimeZone$ \in \reversed $Branch$}{

    \ForEach{$Signal$ \in $TimeZone$}{
      \If{$Signal$ \in $MissingSignals$}{
        $TimeZone$.\createConnectionTo{$Signal$}

        \If{$Signal$.\isAvailable}{
          $MissingSignals \mathrel{-}= Signal$\;
        }
        \Else{
          $MissingSignals.\updateTarget{Signal}$\label{algo:backward:direct:updateTarget}
        }
      }
      \ElseIf{\not $Signal$.\isAvailable}{
        $MissingSignals \mathrel{+}= Signal$
      }
    }
  }

  \caption{Backward Direct propagation:\newline \emph{Requesting unavailable signals upstream}}
  \label{algo:backward:direct}
\end{algorithm}

\paragraph{Direct Backward Propagation}
Algorithm~\ref{algo:backward:direct} details an improved backward strategy which -- based on missing relations -- directly requests connections from the closest TimeZone whereas the missing signal is available.
There are two upstream availability cases: either signal declaration or signal usage.
A key difference compared to the previous strategy is the suppression of successive local placeholder mechanism for signal propagation in favor of direct connections.
Figure~\ref{fig:reso:direct} highlights this difference, as the signal \chisel{e} no longer appears in \emph{addA}, \emph{addB} and \emph{merged\_mul\_mul} TimeZones.
Moreover, the missing relation between \emph{\_init\_} and \emph{xor} TimeZones is replaced by a concrete relation (plain purple arrow) with a latency of 2.

This third and final iteration of resolution algorithm not only solve the abovementioned limitations, but also open an even wider parameterization paradigm.
Although the 3 algorithms all provide perfectly valid synchronization, this last algorithm is the most efficient to illustrate the framework parameterization capabilities.
To that extend and for the sake of conciseness, it is the only algorithm considered for the industrial use-case in Section~\ref{sec:xp:indus}.

\subsection{Parameterization of Resolution Strategies}
\paragraph{Hardware Primitive Selection}
Implementation of direct relations between two TimeZones, as in the latest algorithm, raises the issue of which hardware primitive to pick for a given delay.
While some correspondences are obvious, some others offer an interesting parameterization potential to finely tune the architecture depending on resource availability.
As example, we propose a parameterization based on two integer parameters, $\mathit{depthThreshold}$ and $\mathit{widthThreshold}$, respectively referring to the latency of the path to be implemented and the width of the signal to be propagated.
The following decision table is then used to select which primitive to implement:
\begin{center}
  \small
  \begin{tabular}{@{}c|c|c@{}}
  \toprule
                                   & \multicolumn{2}{c}{Signal width $w$} \\
  Latency $l$                      & $w < \mathit{widthThreshold}$ & $w \geq \mathit{widthThreshold}$ \\
  \midrule
  $l = 0$                          &   \multicolumn{2}{c}{Wire}                                       \\
  $l = 1$                          &   \multicolumn{2}{c}{Register}                                   \\
  $l < \mathit{depthThreshold}$    &   \multicolumn{2}{c}{Successive registers}                       \\
  \cmidrule(lr){1-3}
  $l \geq \mathit{depthThreshold}$ &   Successive registers       &     FIFO                          \\

  \bottomrule
  \end{tabular}
\end{center}
Practical implementation of the FIFO-based primitive is further detailed in Section~\ref{sec:xp:fifo:explo}. 
From this parameterization, we develop the previous \emph{Direct Backward} strategy as follows:
\begin{description}[itemsep=0.2em]
  \item[DirectAuto] Infinite thresholds and let the synthesizer infer shift-registers from registers.
  \item[DirectForceReg] Infinite thresholds and prevent the synthesizer from inferring shift-registers.
  \item[DirectForceSRL] Infinite thresholds and force the synthesizer to infer shift-registers.
  \item[DirectFIFO:D:W] Thresholds $D$ and $W$ and let the synthesizer infer shift-registers.
\end{description}
The results below select a single couple of thresholds (3:3) for \emph{DirectFIFO} strategy to exhibit the ability to influence the behavior of synthesizers for the running pipeline example. 
Such a basic approach can then be further adjusted to better match the missing relations of a given design. 
After inspecting the relations of a complex design, Section~\ref{sec:xp:fifo:explo} demonstrates the flexibility of the framework to explore thresholds and provide a wide range of architectural parameterization.

\paragraph{Uncovered Architectural Parameterization Opportunities}
The three signal propagation algorithms described until this point exclusively focus on propagating individual signals, however many other parameters could be considered.
For the sake of example, a potential parameterization lies in the management of multiple signals following an identical relation.
Their propagation can either be implemented as a single concatenated group of bits (e.g. one large FIFO) or as individual signals (e.g. one FIFO per signal).
Configuring the selection of one method or the other based on signal depth and/or width will yield various synthesis resource usage, which can be used as a mean to fine-tune a design to fit a given target.
Thanks to the methodology presented on Figure~\ref{fig:api}, many distinct implementations of the very same pipeline description can be generated with target-dependent strategies.

\subsection{Comparison of Model Resolution Algorithms}
Table~\ref{tbl:result:running:ex} presents the impact of architectural parameterization on the pipeline described in Listing~\ref{lst:code:running:chisel}.
The five selected strategies, and in particular the four declensions of the \emph{Direct} strategy have been influenced by the respective behaviors of \emph{Vivado (Xilinx)} and \emph{Quartus (Intel)} synthesizers.
In Xilinx FPGA synthesis, Shift Register LUT (SRL) primitives are usually inferred for a signal propagation across at least three register stages sharing the same write-enable signal~\cite{xilinx2018vivado}.
On the contrary, \emph{Quartus} does not automatically infer such SRLs which are not even a base primitive of the Stratix architecture.
For this very reason, to highlight differences between strategies even with a very minimal example, the \chisel{xor} stage is here specified with a delay of $2$ (equivalent to $2$ successive registers) instead of $0$ (combinational wire) as originally specified in Listing~\ref{lst:code:running:chisel}.
The total resulting latency of the circuit becomes $4$ cycles instead of $2$, hence enabling inference of SRLs.
To contradict default behavior of Xilinx synthesizer we leverage the \verilog{(* shreg_extract = "no" *)} Verilog attribute in the \emph{DirectForceReg} strategy.
On the opposite, we leverage the explicit \emph{megafunction} \verilog{altshift_taps} to force Quartus to implement a RAM-based shift-register based on \emph{Memory ALUT} primitives.

\newcommand{\g}[1]{\textbf{#1}}

\begin{table}
  \caption{Resource usage after synthesis: comparison of five resolution strategies, with and without Ready/Valid handshakes. \emph{LLUT: Logic LUT / Combinational ALUT; MLUT: LUTRAM / Memory ALUT; FF: Flip-flop (register)}}
  \label{tbl:result:running:ex}
\begin{center}
  \small
  \begin{tabular}{@{}cl|ccccc|cccc@{}}
  \toprule
  \multicolumn{2}{r}{}          &   \multicolumn{5}{c}{Vivado (Xilinx VU9P)}         & \multicolumn{4}{c}{Quartus (Intel Stratix V)} \\
  \cmidrule(lr){3-7}
  \cmidrule(lr){8-11}
  Protocol & Strategy                  & $\sum$ LUT & LLUT & MLUT &  SRL  & FF      & $\sum$ LUT & LLUT  & MLUT   & FF\\
  \midrule
  \multirow[c]{5}{*}{\shortstack{None \\\footnotesize(\chisel{RawIO})}}
   &  Peer-to-Peer         & 289      & 225    & 0      & \g{64} & 563     &  286      & 286    & 0      & \g{640} \\
   &  DirectAuto           & 289      & 225    & 0      & \g{64} & 563     &  286      & 286    & 0      & \g{640} \\
   &  DirectForceSRL       & 289      & 225    & 0      & \g{64} & 563     &  358      & 294    & \g{64} & 456     \\
   &  DirectForceReg       & 225      & 225    & 0      & 0      & \g{691} &  286      & 286    & 0      & \g{640} \\
   &  DirectFIFO:3:3       & 277      & 237    & \g{40} & 0      & 445     &  431      & 367    & \g{64} & 563     \\
  \midrule
  \multirow[c]{5}{*}{\shortstack{Ready/Valid \\\footnotesize(\chisel{ReadyValidIO})}}
   &  Peer-to-Peer         & 290      & 225    & 0      & \g{65} & 578     &  286      & 286    & 0      & \g{644} \\
   &  DirectAuto           & 290      & 225    & 0      & \g{65} & 578     &  286      & 286    & 0      & \g{644} \\
   &  DirectForceSRL       & 290      & 225    & 0      & \g{65} & 578     &  358      & 294    & \g{64} &     460 \\
   &  DirectForceReg       & 225      & 225    & 0      & 0      & \g{708} &  286      & 286    & 0      & \g{644} \\
   &  DirectFIFO:3:3       & 276      & 235    & \g{40} & \g{1}  & 456     &  424      & 360    & \g{64} &     553 \\

  \bottomrule
  \end{tabular}
\end{center}

\end{table}

The first set of rows corresponds to the absence of handshake between stages, no additional protocol signal is generated and connected. 
The second set, implements a validity signal propagated along the pipeline from the input as well as a common backpressure signal driven by the output.
The backpressure is here applied uniformly across the entire pipeline because there is only one output and the pipeline is already statically scheduled at merging points.
The following paragraphs review the synthesis results obtained with these various strategies and protocols.

\paragraph{Equivalent Resource Usage}
On Xilinx side, for each protocol implementation, \emph{Peer-to-peer}, \emph{DirectAuto} and \emph{DirectForceSRL} strategies exhibit identical resource usage.
This is expected due the systematic inference of successive registers as SRLs, unless it is explicitly avoided as in \emph{DirectForceReg} strategy. 
A contrasting behavior can be observed with \emph{Quartus} synthesizer, which yield identical resource usage for \emph{Peer-to-peer}, \emph{DirectAuto} and \emph{DirectForceSRL} strategies.
No shift-registers are inferred, unless explicitly instantiated as in \emph{DirectForceSRL} strategy.
These results also show that the \emph{Peer-to-peer} and \emph{DirectAuto} strategies are equivalent.
As the \emph{Direct} strategy is much more configurable and flexible, it will be the only one considered in the industrial experimentation.

\paragraph{Impact of Architectural Parameterization}
The impact of \emph{Direct FIFO} strategy appears clearly with the use of memories (LUTRAMs primitives) instead of registers (FFs) or SRLs.
Some control signals are always required to read and write data to a FIFO.
In presence of Ready/Valid handshake signals, control signals of the FIFO are easily driven.
However, in order to implement the FIFO strategy without protocol, an additional piece of hardware must be inserted to guarantee a constant latency through the FIFO.
A counter, initially reset to the expected latency through the FIFO, is decremented at each cycle until it reaches and then maintains the value $0$.
The back-pressure signal, i.e. read enable, of the FIFO is activated only if the counter is equal to $0$.
Compared to a register-based solution, this additional mechanism has a noticeable impact on resource usage due to both FIFO and counter overheads.
In particular, this explains why the \emph{DirectFIFO} strategy requires less resource with ready/valid protocol rather than without, which is a surprising fact at first glance.

By forcing usage of memory primitives in register-oriented context, this convoluted strategy appears to be useful when dealing with local congestion issues on complex design.
Local congestion is indeed due to an unbalanced usage of resource, e.g. an area with a very high demand for registers whereas the LUTs remain barely used.
Leveraging such an enigmatic strategy can reduce the register congestion by instead implementing the exact same signal propagation with LUTRAMs.
Further exploration of the parameterization of this strategy is detailed in Section~\ref{sec:xp:fifo:explo}.

\paragraph{Performance}
The throughput and latency of the resulting pipeline are intended to be guaranteed by design.
The resulting circuit follows exactly the design intent expressed by the designer without any alteration of the stages and explicit delays between TimeZones.
The resolution process enforced here first precisely balances all concurrent paths---merge resolution---then propagate the required signals with exact latencies.
Then protocol signals are implemented and connected as part of the fully synchronous circuit.
Following this pattern, there are no compromises on performance: FIFOs are always dimensioned for the worst case, as this is a requirement for high-throughput network applications.
As a side note, some applicative contexts, which targets average rather than worst-case performance, could reduce the depth of delaying FIFOs based on expected duty cycle.
Implementing such strategies would provide a trade-off between resource footprint and performance.

As a conclusion, this example demonstrates the ability to experiment a wide range of architectural parameterization, yielding very different resource usage, while preserving both original functionality and performance.
The next section develops the use of \emph{PAF} framework at larger scale.


\section{Industrial use-case}
\label{sec:xp:indus}

This section details an industrial application of \emph{PAF} aiming at validating its relevance in high-speed networking application context, considering the two following criteria:
\begin{description}
  \item[Architectural Parameterization] Choices of generation strategies lead to distinct and overall predictable resource usage,
  \item[Zero-cost Pipeline Abstraction] Compared to an equivalent exhaustive description, a circuit generated with the framework exhibits similar resource usage. Strictly tied to the pipeline description, latency and throughput are identical.

\end{description}

\subsection{Applicative Context}
\begin{figure}[tbp]
  \hspace*{0.30em}
  \centerline{
    \begin{subfigure}{0.64\textwidth}
      \includegraphics[width=0.96\textwidth]{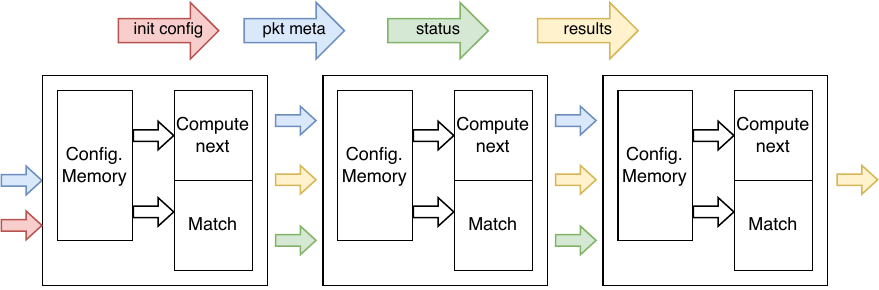}
      \caption{Conceptual design of the classifier illustrated on 3 stages}
      \label{fig:classifier:base}
    \end{subfigure}
    }
  \hspace*{0.30em}
  \centerline{
  \begin{subfigure}{0.64\textwidth}
    \includegraphics[width=0.96\textwidth]{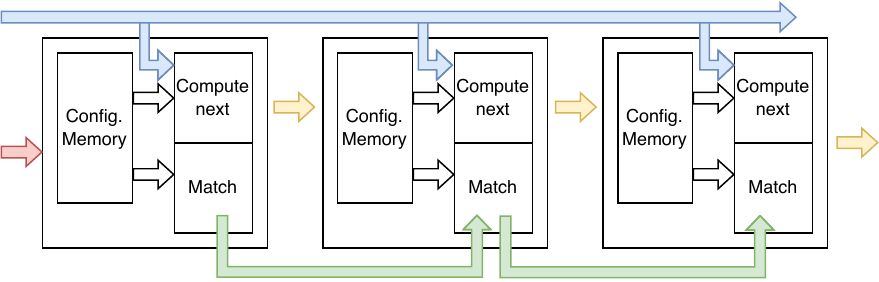}
    \caption{Actual implemented architecture of the classifier}
    \label{fig:classifier:verilog}
  \end{subfigure}
  }
  \caption{Classifier design}
  \label{fig:classifier}
  \vspace*{-1em}
\end{figure}

Tree-based packet classifiers are commonly used as building block for network applications, in order to group packets by categories, mostly depending on their protocol, source and destination addresses.
A packet classifier enables a network appliance to define a flow as a sequence of packets associated to the same label.
This information is then used by downstream processing blocks to process the packets, e.g. dropping them if it is deemed necessary.
Many algorithms and associated architectures can be used to implement packet classifiers, ranging from linear search to geometric algorithms to advanced heuristics~\cite{gupta2001algorithms, taylor2005survey}.
As always in hardware design and implementation, configurability and performance induce considerable resource costs, and a balance must be found between these parameters to provide efficient packet classifiers.
Using a tree approach such as the one proposed in HyperCuts~\cite{singh2003packet}, classification rules can be compressed to keep the implementation affordable in resources, while guaranteeing a one-packet-per-cycle throughput.
The tree structure requires strong synchronization between packet metadata, internal state and results, which makes it an interesting application for our framework.
Figure~\ref{fig:classifier:base} illustrates its base principle: a succession of configurable processing stages, each corresponding to a level in the tree.
Only three stages are depicted here, however the number of stages must be fully configurable and depends on the maximum number of rules expected for a given application.
At each stage, based on the result of the previous stage, the current configuration is retrieved from a memory.
This configuration is then combined with packet metadata and internal tree state to produce 
\begin{enumerate*}
  \item the updated tree state, 
  \item the classification result, and 
  \item the address of the configuration for the next stage.
\end{enumerate*}

In a network context, a network packet metadata usually includes the so-called \emph{5-tuples}: source and destination IP address, protocol, and source and destination ports.
For an IPv6-enabled network device, with $128$-bit IP addresses, $8$-bit protocol identifier, and $16$-bit ports, it results in a bare minimum of a $296$-bit data bus to be distributed to each stage with proper synchronization.
In practice, with an experimentally reasonable number of $40$ stages and $6$-cycle latency per stage ($2$ for memory access, $4$ for computations), it requires already more than $70$k registers only to forward and provide metadata to each stage.
The impact on resource usage of the actual implementation of this signal forwarding across the design is considerable and such architectural choices become critical to eventually meet timing requirements.

The original implementation of this classifier has been written in SystemVerilog and currently targets both Intel and Xilinx/AMD FPGAs, respectively \emph{Stratix V} and \emph{Ultrascale Plus} architectures~\cite{horrein2021method}.
As the number of registers in an ideal implementation, in which data is forwarded along with computations, is too high to allow efficient placement and routing of the design, we instead leverage vendor-provided shift registers, which comes with their own capabilities and limitations.
Xilinx provides LUT-based SRL primitives, able to manage up to $32$-bit long and $1$-bit wide shift registers.
Its toolchain, Vivado, is able to efficiently infer shift registers as long as they do not cross a hierarchy boundary.
On the contrary, Intel toolchain, Quartus, avoids shift register inference in most cases because it uses M20K or MLAB memory blocks, which are a rather rare and precious resource in the FPGA, often harder to route.
To force the use of appropriate vendor primitive, we had to manually extract data forwarding parts of the design, and integrate them directly in the top level module, which breaks the \emph{black-box} paradigm and requires spreading information on internal pipelines lengths.

Based on the theoretical architecture presented in Figure~\ref{fig:classifier:base}, Figure~\ref{fig:classifier:verilog} depicts the \emph{ad-hoc} implementation required to set up this explicit signal forward strategy.
The three different streams of data are passed along the stages with their own explicit forward implementations:
\begin{description}
  \item[Packet Metadata] Read once at the beginning of the stage and not modified internally.
  A tapped shift-register is effective here, to forward metadata aside the pipeline, and supply them synchronously to each stage, after the configuration retrieval.
  \item[Compute Results] Forwarded within the main pipeline, as they are used and modified all along the stages.
    Proper split with computation here is difficult, and can only be done locally in the design, e.g. in parallel with memory access.
  \item[Tree State] Modified only once during the compute stage, after the configuration retrieval.
    This modification is simple and done in a combinational way, permitting the implementation of a separate shift-register-based bypass of the depth of a complete stage latency.
\end{description}

In the end, modularity of the original architecture is greatly restricted with this optimized implementation, which is hard to maintain and upgrade.
For example, adding or removing a register within a compute block requires recomputing and propagating the latency change across the entire pipeline, in order to ensure that data remains properly synchronized.

\subsection{Re-implementation with the Framework}
To experiment with our framework on this classifier, the original Verilog version was first translated to Chisel, using \emph{sv2chisel}~\cite{bruant2020system}.
To avoid any regression, it begins with a word-for-word translation and validation of this initial translated version against the existing test-benches.
After some manual adjustments, the translated version exhibits similar resource usage and ability to match timing in several network applications.
From then on, the validated version is referred as \emph{baseline}, and can be confidently upgraded with advanced features, such as the pipeline automation offered by our framework.

\begin{figure}[ht]
  \begin{center}
    \vspace*{-0.5em}
    \includegraphics[trim=43 270 45 90,clip, width=\textwidth]{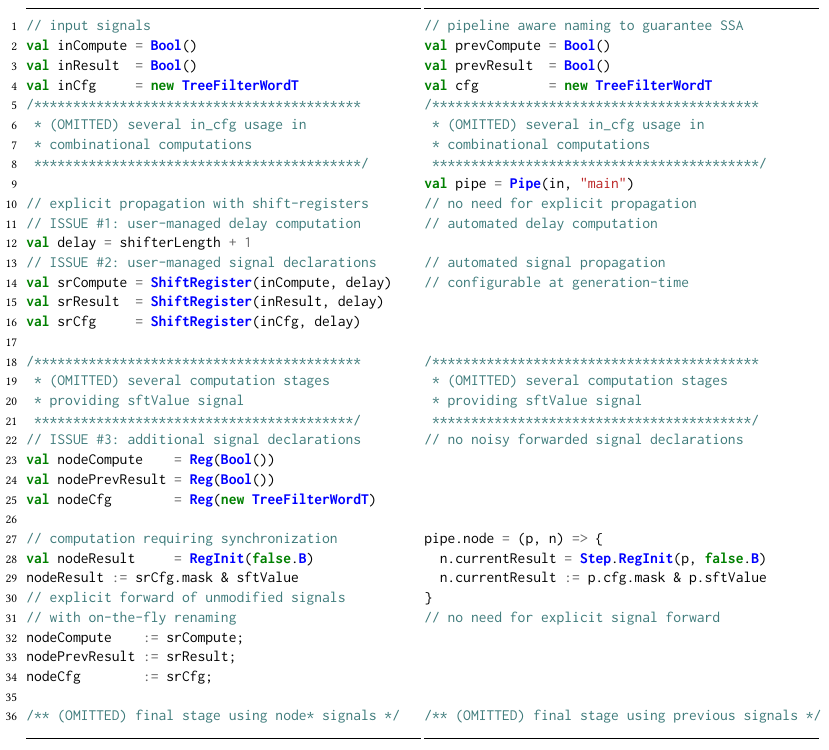}

    \begin{minipage}[t]{0.48\textwidth}
      \vspace*{-1.5em}
        \captionof{listing}{Vanilla Chisel with \\explicit signal propagations}
\label{lst:tf:vanilla:bypass}
    \end{minipage}
    \begin{minipage}[t]{0.46\textwidth}
      \vspace*{-1.5em}
      \captionof{listing}{Pipeline-oriented Chisel version with implicit signal propagations}
\label{lst:tf:chisel:framework:bypass}
    \end{minipage}
  \end{center}

\end{figure}

Listing~\ref{lst:tf:vanilla:bypass} presents an explicit signal propagation to match the synchronization requirements, as originally implemented in the Vanilla Chisel version.
This snippet illustrates several pipeline stages:
\begin{description}[itemsep=0.2em]
  \item[Input stage] l.$1$--$4$, \emph{Only three signals considered here for conciseness}
  \item[Initial combinational stage] l.$5$--$8$, \emph{(omitted)}
  \item[Explicit signal propagation] l.$10$--$19$, \emph{Shift registers synchronizing inputs with computations}
  \item[Intermediate computation stages] l.$21$--$24$, \emph{(omitted)}
  \item[Node stage] l.$26$--$39$, \emph{Requiring synchronization of previous computations with inputs}
  \item[Final stage] l.$40$, \emph{(omitted) Requiring signals from several previous stages}, line $40$ 
\end{description}
The explicit signal propagation used in this implementation raises several issues, detailed as comments in the listing.
The root cause for all these issues is the full responsibility of the developer to manage all details of the synchronization, which results in an error-prone and verbose description. 

To address all these issues at once, Listing~\ref{lst:tf:chisel:framework:bypass} illustrates the equivalent pipeline-oriented description of the same processing stages.
Based on the explicit \chisel{Pipe} object, this implementation natively solves the synchronization requirements between computation stages, without requiring explicit signal propagation.
The PAF-based version is focused on actual processing operation, and freed of noisy statements whose sole purpose is to guarantee proper synchronizations.
Applied to the complete module, it incidentally reduces the amount of lines of code by about $12\%$.
Omitted signal propagations are then programmatically implemented, allowing elaboration-time parameterization and experiments with various synchronization strategies.
Unlike the original implementation, the resulting pipeline description remains quite similar to the initial design in Figure~\ref{fig:classifier:base}, preserving its flexibility.
As a result, exploration and optimization is made easier: modification of the processing part can be done without taking into account data forwarding, while improvements of the data forwarding part can be done by using appropriate strategies in the framework.

Based on its internal pipeline-oriented abstraction, the framework is also able to provide an enriched representation of pipeline stages and theirs signals, focused on synchronization requirements, as previously introduced in Figures~\ref{fig:reso:all} and~\ref{fig:reso:nores}.
The generated representation is very valuable to the designers to track the declaration and usages of the signals across the pipeline.

\subsection{Results}
With its simple composition and configuration strategies, the framework intends to provide a wide diversity of synthesis results, depending on the elaboration-time parameterization of the pipeline. 
The results presented in this section target two distinct FPGA architectures which are currently deployed in production at OVHcloud and implement the same device, except for performance:
\nobreak
\begin{itemize}[itemsep=0.1em]
  \item Xilinx \emph{Virtex Ultrascale+} at \MHz{200}, for a \Gbps{400} throughput. \emph{(Vivado 2018.2)}
  \item Altera/Intel \emph{Stratix V} at \MHz{150}, for a \Gbps{40} throughput. \emph{(Quartus 15.1)}
\end{itemize}
In accordance with the original SystemVerilog and vanilla Chisel implementations, this pipeline-oriented version is here generated without protocol signaling.

\begin{table}[ht]
\caption{Resource usage comparison for a 52-stage packet classifier on Xilinx Ultrascale+}
\label{table:tf:rsc:xilinx}
\begin{center}
  \small
  \begin{tabular}{@{}rcccccccc@{}}
  \toprule

  &             $\sum$ LUT &    LLUT    & LUTRAM &  SRL  &   FF  & RAMB36 & RAMB18 & URAM \\
  \midrule

  Baseline      &   90,316 &     59,266 &     0 & 31,050     &  63,988 &     20 &    190 &   92   \\
  \midrule
  
  DirectAuto    &   90,294 &     59,199 &     0 & \g{31,095} &  64,128 &     20 &    190 &   92  \\ 
                &      -22 &        -67 &     0 &    +45     &    +140 &        &        &       \\
                &   -0.02\% &   -0.11\% &     0 & +0.14\%    & +0.22\% &        &        &       \\
  \midrule
  DirectForceReg  &   57,900 &     57,842 &       0 &   58   & \g{258,860} &     20 &    190 &   92 \\
  DirectFIFO:6:16 &   86,391 &     64,216 &  \g{15,648} &  6,527 &  42,079 &     20 &    190 &   92 \\

  \bottomrule
  \end{tabular}
\end{center}
\end{table}

Table~\ref{table:tf:rsc:xilinx} presents resource usage of a fully realistic 52 stage classifier, as synthesized for the Xilinx \emph{Virtex Ultrascale+} FPGA.
Xilinx toolchain reports three categories of LUT: Logical LUT (LLUT), LUTRAM and Shift register LUT (SRL).
The last two are internally sharing the same \emph{Memory LUTs} resource.
The sum of these three primitives, reported in the leftmost column, is a relevant comparison criterion between resource usage.
Indeed, Xilinx suggests maintaining the ratio between LUTs and FFs around 1 for best placement and routing performance.
%
With regard to our two criteria, these results perfectly fulfil our expectations.
Leveraging Vivado SRL inference behavior, the \emph{DirectAuto} strategy allows us to retrieve an almost identical resource usage compared to original implementation, validating the \emph{zero-cost abstraction} of PAF.
Moreover, both other strategies report very different resource usage.
The \emph{DirectForceReg} strategy is fully based on registers (FF primitive) and considerably reduces the overall LUT usage.
The \emph{DirectFIFO:6:16} strategy, balances the memory LUT usage between SRL and LUTRAM primitives, incidentally reducing register usage.
The choice of the depth and width threshold is purely arbitrary here and is further explored in Section~\ref{sec:xp:fifo:explo}.

\begin{table}[ht]
\caption{Resource usage comparison for a 52-stage packet classifier on Intel Stratix V}
\label{table:tf:rsc:intel}
\begin{center}
  \small
  \begin{tabular}{@{}rccccc|c@{}}
  \toprule
                 & $\sum$ LUT  &  MLUT       &      LLUT &     FF  &    M20k &  ALM    \\
  \midrule
Baseline        &    73,877     &    16,090  &   57,787  & 22,556   &    447  &     51,331.1 \\
\midrule
DirectForceSRL  &    73,604     & \g{15,140} &   58,464  & 23,043   &  447    &   51,525.2 \\
                &      -273     &      -950  &     +677  &   +487   &   =     &     +194.1 \\
                &    -0.4\%     &    -5.9\%  &   +1.2\%  &  +2.2\%  &   =     &    +0.4\% \\
\midrule
DirectAuto      &     43,823    &    80      &   43,743  & 73,281   & \g{1,004} &  \g{33,430.9} \\
DirectFIFO:6:16 &     74,169    & 2,080      &   72,089  & 45,396   & \g{1,116} &   46,770.2 \\

  \bottomrule
  \end{tabular}
\end{center}
\end{table}
Similarly, Table~\ref{table:tf:rsc:intel} presents the resource usage of a slightly differently configured 52-stage classifier fitted on an \emph{Intel/Altera Stratix V} FPGA.
The rightmost column of the table reports an estimate number of logic elements (ALMs) actually used on the FPGA.
A lower value means a reduced circuit footprint and more efficient resource packing.
The \emph{DirectAuto} strategy minimize this count, at the cost of increased M20K utilization.
Despite its higher ALM count, the baseline implementation has been carefully chosen because it appears easier to route within complete network appliance design.
The \emph{DirectForceSRL} strategy achieves close resource usage to this baseline implementation, in particular preserving the amount of total ALM needed, even if the balance of LUTs and registers is slightly different.
This strategy has the property to prevent the synthesizer to infer M20K-based SRL, keeping the exact same count as the baseline.
Such behavior is highly desired here to preserve these rare resources for other usage in the complete network appliance design.
Automated inference of large SRL, using M20K (block memory) primitives is indeed demonstrated by the \emph{DirectAuto} strategy.
Finally, the \emph{DirectFIFO:6:16} strategy illustrates another utilization balance, also based on M20K primitives but trading registers for LUTs compared to the \emph{DirectAuto} strategy.

As a conclusion, synthesis results perfectly meet our expectations on both targets and regarding both evaluation criteria.
PAF provides a pipeline-oriented abstraction at zero resource utilization cost and offers extensive architectural parameterization.
This successful validation of architectural parameterization capabilities leads to an additional question: how to find the best parameterization for a given use-case?
This intricate question cannot find a universal answer as it is highly dependent on both target and surrounding integration of the current design.
However, the next section details how PAF aim at providing the tools to help the designer to review a wide range of potential configurations, in a matter of minutes.

\section{Analysis and Automation Tools for Architectural Parameterization}
\label{sec:xp:fifo:explo}

The previous sections detailed PAF's ability to expose numerous architectural parameters and how these parameters yield direct impact on the resulting resource usage.
This powerful feature provides a lot of parameterization power to designers, but it also assigns them to select a parameter set matching the needs of the design.
Taking the right decision requires a broad overview of the design and its content.
Thanks to the synchronization model, PAF is able to:
\begin{enumerate*}
  \item report the missing relations, and group them by resolution parameters \emph{(e.g. depth and width)},
  \item generate all the parameterized version of the design.
\end{enumerate*}
These designs can then be fed to synthesis tools to select the most suited parameter set, for a given target and in a given context.

The present experimentation intends to detail the fine-grained parameterization of \emph{DirectFIFO} strategy in the context of the industrial use-case introduced in Section~\ref{sec:xp:indus}.
To better explain the parameterization abilities of this strategy, based on the depth and width of the missing relation, we first detail its implementation.

\subsection{FIFO-based Latency-constant Signal Propagations}

FIFOs (queues) are tailored to act as variable-latency paths, as they provide two independent interfaces, \emph{write} (\emph{enqueue}) and \emph{read} \emph{(dequeue)} interfaces.
In that regard, they are often used as buffers to compensate or synchronize with other variable-latency paths, e.g. to accept a back-pressure at the output of a memory which itself does not accept such a back-pressure.
To play this role, such variable latency paths require protocol signaling to synchronize with one another.
In the case of FIFO, the \emph{write} interface is connected to the upstream protocol signals and the \emph{read} interface is connected to the downstream protocol signals as follows:
%
\begin{lstlisting}
  upstream.ready := !fifo.full
  fifo.write_enable := upstream.valid

  downstream.valid := !fifo.empty
  fifo.read_enable := downstream.ready
\end{lstlisting}
From then on, if we attempt to remove the protocol signaling around this FIFO, considering that the circuit is uniformly exempt of back-pressure and its data is always valid, the following connections are observed:

\begin{lstlisting}[language=scala]
  fifo.write_enable := true.B
  fifo.read_enable := true.B
\end{lstlisting}
As soon as the circuit powers up\footnotemark, this results in a continuous flow of data through the FIFO with a $1$-cycle latency\footnotemark between its \emph{write} and \emph{read} interfaces, regardless of the depth of the FIFO.
To create a constant $N$-cycle latency path with such FIFO, we need to delay the \chisel{fifo.read_enable} activation by $N$ cycles.
\footnotetext[\numexpr\thefootnote-1]{Or is reset, in such case, \chisel{true.B} should be read as \chisel{!reset} for an active-high reset.}%
\footnotetext[\numexpr\thefootnote-0]{$1$-cycle latency is achieved with simple memory implementation. Without loss of generality, this latency might be increased as long as this higher latency remains below the signal propagation depth.}%
Such behavior can be obtained with a register \chisel{read_start}, initially reset to the value $N$ and decremented at each cycles until it reaches and maintains the value $0$.
The read interface is then connected as follows:
\begin{lstlisting}[language=scala]
  fifo.read_enable := read_start === 0.U
\end{lstlisting}
This results in an equivalent behavior to a shift-register of $N$-cycle, but it is described based on a FIFO architecture which uses memory block primitives.

Based on this implementation, the \emph{DirectFIFO:D:W} strategy introduces two configurable thresholds above which a given signal propagation will be implemented with a FIFO rather than shift registers.
\begin{enumerate}[itemsep=0.2em]
  \item $D$: Minimum signal propagation depth threshold
  \item $W$: Minimum signal width threshold
\end{enumerate}
Leveraging minimum thresholds is a first naive approach which is justified by the non-linear resource cost of FIFOs in terms of both depth and width.
FIFOs are indeed based on block memory primitives whose configuration is restricted to a set of \emph{(depth, width)} couples depending on the FPGA architectures.
Fully exploiting the parameterization of this strategy first requires an analysis of the signal propagation depths and widths through the design.

\subsection{Exhibiting and Exploiting Architectural Parameterization}

In addition to delivering zero-cost pipeline-oriented abstraction and architectural parameterization, PAF also provides tools to efficiently analyze the design.
In particular, it is able to report an overview of the depth and width of signal propagations to be implemented and how they have been implemented by the current strategy.
Table~\ref{table:tf:depth:width:usage} details the occurrences of the respective \emph{(depth, width)} couples of all direct signal propagations being implemented by the framework, with depth superior or equal to $3$. 
The submodule \emph{TfCompute}, appearing in each stage of the classifier with the exact same parameterization is elaborated only once as an independent pipeline, and is then instantiated $97$ times in the final circuit.

\begin{table}
\caption{Distribution of direct signal propagations to be implemented in the design, as reported by PAF after pipeline elaboration}
\label{table:tf:depth:width:usage}
\begin{center}
\small
\begin{tabular}{@{}rcc|cc|cc@{}}
\toprule
& \multicolumn{2}{c|}{\textbf{Specification}}    & \multicolumn{3}{c}{\textbf{Occurrences}} \\ 
&   Depth    &  Width      &  TfCompute &    TfTop  &   Total \\ 
\midrule
& 4          &   64        &  $1 \times 97$  &           &    97   \\  
& 5          &   1         &  $2 \times 97$  &           &   194   \\
\midrule
& 6          &   1         &                 &      45   &   45    \\ 
& 6          &   12        &                 &      1    &   1     \\ 
& 6          &   264       &                 &      1    &   1     \\ 
\midrule
& 8          &   6         &                 &      6    &   6     \\ 
& 8          &   12        &                 &      90   &   90    \\ 
& 8          &   62        &                 &      96   &   96    \\ 
& 8          &   132       &                 &      6    &   6     \\ 
& 8          &   246       &                 &      51   &   51    \\ 
& 8          &   264       &                 &      45   &   45    \\ 
\midrule
& 419        &   1         &                 &      1    &   1     \\

\bottomrule
\end{tabular}
\end{center}
\end{table}

Based on this report, the relevant \emph{(depth, width)} minimum thresholds can be generated with simple scripts.
These thresholds are then used as parameterization of the \emph{DirectFIFO:D:W} strategy, which produces in a few minutes the $22$ corresponding Chisel-generated Verilog hierarchies.
A few more scripts are then sufficient to launch the corresponding synthesis and report their respective resource usage, unattended and in a matter of hours.

\newcommand{\me}[1]{\multicolumn{6}{r||}{\footnotesize\emph{#1}}}
\newcolumntype{+}{>{\global\let\currentrowstyle\relax}}
\newcolumntype{=}{>{\currentrowstyle}}
\newcommand{\rsty}[1]{\gdef\currentrowstyle{#1}%
    #1\ignorespaces
}

\begin{table}
  \caption{Resource usage for the relevant threshold parameterization}
  \label{table:tf:depth:width:resource:usage}
\begin{center}
\small
\begin{tabular}{@{}+c|=c||=c=c=c|=c=c|=c=c|=c=c@{}}
\toprule
Depth & Width    & LUTRAM   &   SRL  &  LLUT      & $\sum$LUT  &      FF    &   LUT+FF   & LUT/FF  & RAMB18 & URAM \\
\midrule
$\infty$ & $\infty$ & \g{0} & 31,095 &    59,199  &   90,294   &    64,128  &   154,424  &   1.41  &  190  & 92 \\
419  &   1       &      16  & 30,828 & \g{59,156} &   90,000   &    64,635  &   154,635  &\g{1.39} &  190  & 92 \\
\midrule
8    &   264     &   4,288  & 25,229 &    61,657  &   91,174   &    58,943  &   150,117  &   1.55  &  190  & 92 \\
6    &   264     &   4,384  & 25,129 &    61,385  &   90,898   &    58,727  &   149,625  &   1.55  &  190  & 92 \\
8    &   246     &  11,488  & 12,966 &    62,440  &   86,894   &    47,218  &   134,112  &   1.84  &  190  & 92 \\
6    &   246     &  11,584  & 12,836 &    62,354  &   86,774   &    47,181  &   133,955  &   1.84  &  190  & 92 \\
8    &   132     &  11,824  & 12,430 &    62,517  &   86,771   &    46,739  &   133,510  &   1.86  &  190  & 92 \\
6    &   132     &  11,920  & 12,296 &    62,644  &   86,860   &    46,621  &   133,481  &   1.86  &  190  & 92 \\
8    &   62      &  15,552  &  6,662 &    64,148  & \g{86,362} &    42,184  &   128,546  &   2.05  &  190  & 92 \\
6    &   62      &  15,648  &  6,527 &    64,216  &   86,391   & \g{42,079} &\g{128,470} &   2.05  &  190  & 92 \\
4    &   64      &  15,720  &  6,896 &    66,266  &   88,882   &    42,965  &   131,847  &   2.07  & \g{0} & 92 \\
8    &   12      &  16,264  &  6,130 &    65,631  &   88,025   &    42,822  &   130,847  &   2.06  &  190  & 92 \\
8    &   6       &  16,312  &  6,094 &    65,780  &   88,186   &    42,854  &   131,040  &   2.06  &  190  & 92 \\
8    &   1       &  16,328  &  5,835 &    65,732  &   87,895   &    43,380  &   131,275  &   2.03  &  190  & 92 \\
6    &   12      &  16,368  &  5,984 &    65,688  &   88,040   &    42,698  &   130,738  &   2.06  &  190  & 92 \\
6    &   6       &  16,416  &  5,948 &    65,782  &   88,146   &    42,728  &   130,874  &   2.06  &  190  & 92 \\
6    &   1       &  16,520  &  5,891 &    66,204  &   88,615   &    43,161  &   131,776  &   2.05  &  190  & 92 \\
\midrule
5    &   1       &  16,900  &  5,752 &    68,199  &   90,851   &    44,997  &   135,848  &   2.02  &  190  & 92 \\
4    &   62      &  19,448  &  1,130 &    67,819  &   88,397   & \g{38,424} &   126,821  &   2.30  & \g{0} & 92 \\
4    &   12      &  20,168  &    584 &    69,281  &   90,033   &    39,039  &   129,072  &   2.31  & \g{0} & 92 \\
4    &   6       &  20,216  &    548 &    69,410  &   90,174   &    39,073  &   129,247  &   2.31  & \g{0} & 92 \\
4    &   1       &  20,700  &\g{106} &    71,794  &   92,600   &    41,854  &   134,454  &   2.21  & \g{0} & 92 \\
\bottomrule
\end{tabular}
\end{center}
\end{table}

Table~\ref{table:tf:depth:width:resource:usage} presents the result of this quick exploration study after synthesis on the abovementioned Xilinx toolchain.
Results are ordered by ascending count of LUTRAMs whose presence corresponds to FIFOs.
As expected, the resource usage of the infinite thresholds is equal to the resource usage of \emph{DirectAuto} strategy previously reported in Table~\ref{table:tf:rsc:xilinx}.
Similarly, resource usage for the default parameterization \emph{(6, 16)} of the \emph{Direct FIFO} strategy is equal to the \emph{(6, 62)} parameterization as both match the same signal propagations for this design.
Interestingly, this default parameterization is close to the median value in terms of LUTRAM count which confirms the relevance of these arbitrarily selected default thresholds for the current design.
This parameterization is also the one minimizing the cumulated count of registers and LUTs.
Regarding the ratio between LUTs and registers, the original strategy remains the closest to Xilinx guidelines.
An interesting alternative strategy to reduce LUT count could be to force BRAM primitives instead of LUTRAM for FIFO implementation.
However, it is not an option in the current network appliance, as almost all BRAMs are already required by other parts of the design. 

Finally, Table~\ref{table:tf:depth:width:resource:usage} reveals an odd behavior for a \emph{(4,64)} propagation whose output drives a Read-Only Memory (ROM).
When using a FIFO instead of a shift-register to propagate the read address of the ROM, the synthesizer decides to get rid of the BRAM primitive and to include its content within the propagating LUTRAMs.
This BRAM inference and simplification behavior is altered with \emph{Vivado 2018.3} which results in a major difference in the ability to meet timing requirements for some of our larger designs.
Based on the exploration capabilities of the framework, this subtle difference is not only highlighted, but the framework also provides means to compensate for it.
In this precise case, this would be achieved by either using a shift register for this particular path, or leveraging explicit Xilinx BRAM primitives for the FIFO implementation.

With this wide range of architectural parameterization, our pipeline automation framework demonstrates its ability to provide both guidance -- to select the relevant parameterization -- and efficiency -- to generate different pipeline implementations with a considerable impact on resource usage.
After synthesis of the various implementations, their resource usage is widely spread across the design space which provides a great flexibility to the designer.
In particular, the ratios of LUTs against registers can be tuned, which might help the routing step with timing closure.
While raising the level of hardware abstraction to a convenient pipeline-oriented modeling, PAF also increases designers' control over the finest architectural details of their designs.

\section{Conclusion \& Future Works}
\label{sec:conclusion}

In this paper, we have introduced a pipeline design methodology aiming at providing extended architectural parameterization capabilities.
It comes in the form of a fully decoupled API, with on one hand expression of pipeline stages behavior, and on the other hand synchronization strategies and protocol signaling implementations.
We close the gap between these two interfaces with a graph-based synchronization model and by providing, at elaboration-time, fine-grained control to users over the generated design.
By leveraging Chisel HCL to implement our methodology as an automation framework, we have successfully migrated a complex existing pipeline at no additional resource cost while considerably relieving developers' synchronization burden.
Moreover, decoupling protocol signaling and synchronization concerns from the pipeline description highly improves code readability and reusability while advantageously reducing its volume.

We now plan to extend this automation approach to variable latency propagations and multi-level pipelines processing data at various abstraction levels.
Adding support for automated conversion from one level to another would greatly ease interaction between cross-level modules and enable extended architectural parameterization at almost no redesign cost.


\end{document}